\documentclass[aps,prd,letter,superscriptaddress,amsmath,twocolumn,floatfix,showpacs]{revtex4-1}
\newcommand{\beq}{\begin{equation}}
\newcommand{\eeq}{\end{equation}}
\newcommand{\beqn}{\begin{eqnarray}}
\newcommand{\eeqn}{\end{eqnarray}}

\newcommand{\apjl}{Astrophys.\ J.\ Lett.}
\newcommand{\apjs}{Astrophys.\ J.\ Supp.}

\newcommand{\llabel}[1]{\label{#1}}              % DO NOT show equation label
\newcommand{\labeq}[2]{ \begin{equation} \llabel{#1}{#2}
\end{equation}}
\usepackage{graphicx}
\usepackage{epsf}
\usepackage{graphics,epsfig,placeins,subfigure,wrapfig,amssymb,pifont}

\begin{document}
\title{Importance of cooling in triggering the collapse of hypermassive neutron stars}

\author{Vasileios Paschalidis}
\author{Zachariah B. Etienne}
\author{Stuart~L.~Shapiro}
\altaffiliation{Also at Department of Astronomy and NCSA, University of
  Illinois at Urbana-Champaign, Urbana, IL 61801}
\affiliation{Department of Physics, University of Illinois at
  Urbana-Champaign, Urbana, IL 61801}

\begin{abstract}

The inspiral and merger of a binary neutron
star (NSNS) can lead to the formation of a  hypermassive neutron star (HMNS). As
the HMNS loses thermal pressure due to neutrino cooling and/or centrifugal support due
to gravitational wave (GW) emission, and/or magnetic breaking of differential
rotation it will collapse to a black hole. To assess the importance of shock-induced thermal
pressure and cooling, we adopt an idealized equation of state and perform NSNS simulations
in full GR through late inspiral, merger, and HMNS formation, accounting 
for cooling. We show that thermal pressure contributes
significantly to the support of the  HMNS against collapse and that thermal
cooling accelerates its ``delayed'' collapse. Our simulations  
demonstrate explicitly that cooling can induce the catastrophic collapse of a {\it hot} hypermassive 
neutron star formed following the merger of binary neutron stars.
Thus, cooling physics is important to include in NSNS merger calculations to
accurately determine the lifetime of the HMNS remnant and to 
extract information about the NS equation of state, cooling mechanisms, 
bar instabilities and B-fields from the GWs emitted during 
the transient phase prior to BH formation.

\end{abstract}


\maketitle

\section{Introduction}

The inspiral and merger of compact binaries has attracted
considerable attention in recent years for two main reasons. 
First, such systems emit a large flux of gravitational waves (GWs), 
making them among the most promising sources for GWs detectable by 
ground-based laser interferometers such as LIGO \cite{LIGO1,LIGO2}, 
VIRGO \cite{VIRGO1,VIRGO2}, GEO \cite{GEO}, and KAGRA \cite{KAGRA},
as well as by proposed space-based interferometers 
such as eLISA/NGO \cite{NGO} and DECIGO \cite{DECIGO}. Second, black hole --
neutron star (BHNS) and neutron star -- neutron star (NSNS) mergers are 
candidates for the central engines that power the observed short-hard 
gamma ray bursts (sGRBs). 

Extracting physical information about these binaries
from their GWs and their accompanying electromagnetic signals 
may reveal critical details about the equation of state of neutron
star matter and may unveil the nature of the sGRB phenomenon.  
However, interpreting the data
requires careful modeling of these systems in full general relativity
(see \cite{BSBook} for a comprehensive review and references).
Most effort in general relativity to date has focused on modeling black hole--black hole (BHBH) binaries  
(see also \cite{Hinder:2010vn}),
%\cite{Pretorius2005a,Baker2006,Campanelli2006,Brugmann2006a,
%Pretorius2006,Baker2006a,Campanelli2006a,Herrmann2006,Sperhake2006,Pollney2007,Etienne2007a, Scheel2008,Pretorius2007a,
%Gonzalez2007,Campanelli2007,Gonzalez07,Campanelli2007a,Healy,Baker2006b,Hermann07,Koppitz2007,Brugmann2007,Lousto2007, 
%Campanelli2006d,Gonzalez08,Campanelli2010BHBH,Rezzola2010BHBH,Hannam2010BHBH} 
and neutron star--neutron star (NSNS) binaries (see also \cite{BNSlr}),
% \cite{2002PhRvD..65b4016D,2003astro.ph..6481K,2003PhRvD..68h4020S,2004PhRvD..69l4036F,
%2005A&A...431..297B,2005PhRvD..71h4021S,2006PhRvD..73f4027S,
%2008PhRvD..77b4006A,2008PhRvD..78b4012L,2008PhRvD..78h4033B, 2010CQGra..27k4105R,2010PhRvD..81l3016H,2010arXiv1007.1754B}, 
with some recent work on black hole--neutron star binaries 
(see also \cite{st11}), and white dwarf--neutron star 
binaries \cite{Paschalidis:2010dh,PLES2011,Paschalidis:2009zz}.

%\cite{SBS2000,DLSS2004,dlsss06a}

NSNSs are known to exist, which makes NSNS systems particularly attractive to study. 
Theoretical calculations show that NSNS mergers can lead to 
the formation of a hypermassive neutron star. 
A HMNS \cite{HMNS} is a differentially 
rotating NS whose mass exceeds the maximum mass of a uniformly rotating
star \cite{CST1994a,CST1994b}. The latter is about 20\% larger than 
the maximum mass of a nonrotating (spherical) equilibrium star (the TOV limit)
\cite{HMNS}. Typically a HMNS forms following the merger of a NSNS, when the system's total
mass is smaller than some threshold mass $M_{\rm th}$. According to \cite{ST} 
this threshold mass is $M_{\rm th} \approx 1.3-1.35 M_{\rm sph}$, 
where $M_{\rm sph}$ is the TOV limit for the same EOS.  

A HMNS is a transient, quasiequilibrium configuration. It will eventually
undergo ``delayed collapse'' on a secular (dissipative) time scale, 
which may power a sGRB. There are two
distinct routes by which this collapse might be triggered: 
\begin{enumerate} 

\item If the HMNS is primarily centrifugally supported, redistribution
  of angular momentum by viscosity or magnetic fields \cite{DLSS2004,dlsss06a},
  and/or loss of angular momentum by GW emission 
  \cite{SHTA2006} destroys the support provided,
  leading to catastrophic collapse. 

\item If the HMNS is primarily supported by thermal pressure generated by
  shocks during merger, delayed collapse
  may be triggered by the loss via neutrino cooling 
  of thermal energy~\footnote{Note that neutrinos too carry away angular momentum from the system, 
  but according to \cite{Baumgarte:1998sn} neutrino emission is very inefficient in 
  decreasing the angular momentum of a HMNS.}. 
\end{enumerate} 

While catastrophic collapse of a {\it cold} HMNS via viscosity or 
magnetic fields has been demonstrated using fully general relativistic 
calculations \cite{DLSS2004,dlsss06a}, there are no fully general 
relativistic calculations to date that demonstrate explicitly that 
cooling can induce collapse of a {\it hot} HMNS produced following the 
merger of binary neutron stars.

HMNSs formed in NSNS mergers
will always be {\it hot} due to shock heating.
A priori it is not clear which mechanism is 
most important for holding up a  HMNS against collapse: 
{\it centrifugal forces} or {\it thermal pressure}. The answer to this question is
still open and may depend on the nature of the companions (e.g. masses, EOS etc.).

Recent simulations of binary NS mergers that form 
hypermassive NSs seem to point in different directions. 
For example, in \cite{2008PhRvD..78h4033B,Rezzolla:2010fd} an equal-mass NSNS is evolved 
assuming a $\Gamma=2$ equation of state (EOS). It is
shown that angular momentum carried away by gravitational waves
alone can induce the collapse.
Reference \cite{2011PhRvL.107e1102S} also evolves an equal-mass NSNS, 
but with a more realistic, finite temperature, nuclear EOS. They find that the deviation of their 
HMNSs from axisymmetry is so small that GW emission is significantly reduced.
The authors argue that shock heating is sufficiently important that their  HMNSs
are supported by the excess thermal pressure. 

Determining which mechanism controls the lifetime of the remnant
is important because it determines the time 
interval between the NSNS merger and the delayed collapse
-- a time interval that can in principle be measured by
Advanced LIGO/VIRGO. It is the time interval between the end
of the gravitational wave signal due to the inspiral 
and the beginning of the burst signal due to the delayed collapse. If differential rotation support is most important, 
then the time interval is governed by, e.g. the Alfv\'{e}n time scale, 
assuming magnetic braking of differential rotation is most important, or the GW time scale, in 
the case of a rapidly spinning remnant that develops a bar. 
By contrast, if thermal pressure is dominant, then the time scale 
is governed by thermal cooling. 
Therefore, knowing the mechanism driving collapse may place
constraints on seed magnetic field magnitudes, or the existence of bar modes,
or the relevant cooling mechanisms. It could even
place constraints on the temperature of matter, as well as the nuclear EOS. 

To disentangle the effects of thermal support from those of rotational support,
previous studies compared results from NSNS simulations that suppress shocks 
(by enforcing a strictly cold EOS) to those that allow shocks.
If the HMNS remnant lives longer with
shocks than without, then it is tempting to infer
that thermal pressure due to shock heating is chiefly responsible
for supporting the remnant.
However it is not possible to draw such a firm conclusion because shocks,
which act on a hydrodynamical time scale, not only heat the gas,
thereby increasing the total pressure support, but also affect the
matter and angular momentum profiles. Different profiles
can themselves increase the lifetime of a HMNS.

The goal of this paper is to study 
the relative importance of thermal pressure in supporting  HMNSs 
from collapse and demonstrate that cooling can induce 
the catastrophic collapse of a  HMNS formed following the 
merger of binary neutron stars. We accomplish this by
performing a limited set of NSNS simulations in full 
GR through late inspiral, merger, (hot) HMNS formation, and collapse. 
We account for cooling in the HMNS remnant via a covariant cooling scheme we
developed in \cite{PLES2011}. We then compare this HMNS evolution to
a control simulation, in which the cooling mechanism is disabled.

Our simulations model the initial NSNS binary as equal-mass,
irrotational, quasiequilibrium $n=1$ polytropes in a quasicircular
orbit, corresponding to case \mbox{$1.46$-$45$-$\ast$} of
\cite{2008PhRvD..78h4033B}. 

Following the NSNS merger, a quasiequilibrium HMNS forms. We then
continue the evolution of the remnant with and without 
cooling, which we model via an effective local emissivity. For the runs with cooling we choose two
cooling time scales. We find that, independent of the cooling time scale chosen, 
the HMNS collapses and forms a BH within a few cooling time scales. 

Our simulations suggest that shock-induced thermal 
pressure is a significant source of support against
gravitational collapse, even in the case of polytropic NSs
and demonstrate explicitly that cooling can induce 
the catastrophic collapse of a  HMNS. Estimating the
temperature of the remnant, we find that a realistic neutrino cooling time scale 
is of order a few $100$ms. Given that our estimated cooling 
time scale is comparable to the angular momentum redistribution/loss time scales due
to either magnetic braking or GWs, our results suggest that accounting
for cooling is a critical ingredient in predicting the lifetime of a  HMNS.
Accordingly, cooling physics must be incorporated in models of binary NS simulations.

The paper is structured as follows. In Sec.~\ref{sec:timescales} 
we review the time scales relevant to HMNSs formed in binary NSNS mergers. 
Sections~\ref{sec:basic_eqns} and~\ref{sec:numerical} summarize the
initial data, basic evolution equations, numerical methods, and
cooling formalism. 
The basic results are presented in Sec.~\ref{sec:results} and
summarized in Sec. \ref{sec:summaryandfuturework}.
Throughout this work, geometrized units are
adopted, where $G = c = 1$, unless otherwise specified.

\section{Time Scales}
\label{sec:timescales}

The relevant time scales in the evolution of a typical HMNS 
formed in NSNS mergers are its rotation period $T$, 
the gravitational wave time scale 
$t_{\rm GW}$, 
the cooling time scale
$t_{\rm cool}$, and Alfv\'{e}n time scale $t_{\rm A}$. 
We provide rough estimates of these time scales in 
this section. 

\subsection{Rotation period}
We express the HMNS angular frequency $\Omega$ as some fraction $\epsilon$
of the break-up angular frequency $\Omega_{\rm ms}$ 
\labeq{}{
\Omega\approx \epsilon\sqrt{\frac{M}{R^3}},
}
where $M$ is the HMNS mass and $R$ its radius. The rotation period of the
HMNS can then be written as 
\labeq{}{\begin{split}
T \equiv & \ \frac{2\pi}{\Omega} =  \frac{2\pi R^{3/2}}{\epsilon M^{1/2}} \\
   \approx & \ 2 \bigg(\frac{\epsilon}{0.5}\bigg)^{-1} \bigg(\frac{R}{20 \rm km}\bigg)^{3/2} \bigg(\frac{M}{2.8 M_{\odot}}\bigg)^{-1/2} \rm ms.
\end{split}
}
For the numerical estimate we have used the values for the mass and radius of a typical HMNS 
remnant.

\subsection{Gravitational wave time scale}
GW emission sets the time scale of angular momentum loss from the system. 
The gravitational wave time scale for a triaxial, incompressible, 
spinning ellipsoid with ellipticity $e$
can be estimated as \cite{Shapiro}
\labeq{tGW}{
\begin{split}
t_{\rm GW} \equiv &\ \frac{J}{dJ/dt} \approx \frac{1}{M R^2 \Omega^4 e^2} = \frac{R^4}{\epsilon^4 e^2 M^3}  \\
      \approx  &\ 200 \bigg(\frac{\epsilon}{0.5}\bigg)^{-4} \bigg(\frac{e}{0.75}\bigg)^{-2} \bigg(\frac{R}{20 \rm km}\bigg)^{4} 
      \bigg(\frac{M}{2.8 M_{\odot}}\bigg)^{-3} \rm ms,
\end{split}
}
where $J \approx M R^2\Omega$ is the HMNS angular momentum and the ellipticity is defined as 
\labeq{}{
e = \frac{a-b}{R}, 
}
where $a$ is the semi-major axis of the HMNS, $b$ the semi-minor axis, and
$R$ is $(a+b)/2$. To estimate the time scale, we
assumed a value for the ellipticity that
corresponds to a plausible bar. Note also that our estimated $t_{\rm GW}$ is
comparable to the GW time scale inferred by direct numerical
simulations in \cite{Rezzolla:2010fd}. 

\subsection{Cooling time scale}

HMNSs are cooled predominantly by emission of neutrinos. 
At densities $\gtrsim 10^{11} \rm g/cm^3$ neutrinos become
trapped \cite{Shapiro}. Therefore, the cooling time scale is set by the time it takes
for the neutrinos to diffuse out of the hot HMNS remnant. The
main sources of opacity are free nucleon scattering and neutrino absorption by nucleons
(since protons and neutrons comprise the bulk of the HMNS). The diffusion time scale
can be estimated as \cite{Rosswog:2003rv}
\labeq{tCool}{
\begin{split}
t_{\rm cool} \approx &\ 3\frac{R^2}{\lambda_n c} 
\end{split}
}
where $\lambda_n$ is the mean free path of the neutrinos given by
\labeq{mfp}{
\lambda_n^{-1} = n \sigma_n 
}
where $n$ is the neutron number density\footnote{Given the low value of $Y_e \approx 0.1$,
i.e., of the mean number of electrons per baryon found in NSNS mergers in \cite{Rosswog:2003rv} in our estimates
here we assume for simplicity that almost all baryons are neutrons.}, $\sigma_n$ is the total interaction cross 
section $\sigma_n = \sigma_{\rm scat}+ \sigma_{\rm abs} $, where the elastic 
scattering and absorption cross sections are respectively given by \cite{Shapiro,Rosswog:2003rv}
\labeq{crosssect}{
\begin{split}
\sigma_{\rm scat} \approx &\ \frac{1}{4}\sigma_0 \bigg(\frac{E_\nu}{m_e c^2}\bigg)^2, \\
\sigma_{\rm abs} \approx &\ 1.42 \sigma_0 \bigg(\frac{E_\nu}{m_e c^2}\bigg)^2,
\end{split}
}
where $\sigma_0 = 1.76\times 10^{-44} \rm cm^2$, $m_e$ is the electron mass, and $E_{\nu}$ the neutrino energy. 
Substituting Eqs. \eqref{mfp} and \eqref{crosssect} in Eq. \eqref{tCool} we find
\labeq{tCool2}{
\begin{split}
t_{\rm cool} \approx &\ \frac{15 M \sigma_0(E_\nu/m_e c^2)^2}{4 \pi m_{\rm n} R c} \\
\approx &\ 400 \bigg(\frac{M}{2.8 M_{\odot}}\bigg)\bigg(\frac{R}{20 \rm km}\bigg)^{-1}\bigg(\frac{E_\nu}{10\rm MeV}\bigg)^2 \rm ms,
\end{split}
}
where $n=\bar\rho/m_n$, with $\bar\rho = 3 M/ 4\pi R^3$ the mean HMNS density, and $m_n$ the mass 
of a neutron. For the numerical estimates above we used typical rms
values for the neutrino energy of order $10$MeV, as found in the simulations of
\cite{Rosswog:2003rv}. Note that for typical neutrino energies of
20MeV found in \cite{2011PhRvL.107e1102S} the neutrino cooling
timescale is $\sim 2$s. Both of these works used approximate neutrino
transfer schemes. We see that obtaining a neutrino cooling time scale
depends on identifying the energy(ies) of typical neutrino(s), which in turn requires
accurate modeling of not only bulk motion but also the microphysics.

\subsection{Alfv\'{e}n time scale}
Magnetic fields set the time scale for the braking of differential rotation
in typical HMNSs. This occurs on the Alfv\'{e}n time scale \cite{dlsss06a},
given by
\labeq{talfven}{
\begin{split}
t_{\rm A} \approx  &\ \frac{R}{v_A} \approx \frac{R\sqrt{4\pi \rho}}{B} \\
     \approx &\ 100 \bigg(\frac{R}{20 \rm km}\bigg)^{-1/2}
     \bigg(\frac{M}{2.8 M_{\odot}}\bigg)^{1/2}\bigg(\frac{B}{10^{15} \rm G}\bigg)^{-1} \rm ms,
\end{split}
}
where $v_A$ is the Alfv\'{e}n velocity, and where a strong but dynamically 
unimportant interior magnetic field has been assumed for the numerical estimate. 
While little is known about the strength of NS {\it interior} magnetic fields, 
the value appearing in \eqref{talfven} is consistent with magnetars models \cite{Magnetar}.
In addition, NSNS simulations indicate that magnetic instabilities can amplify
interior B-fields from $\sim 10^{12}$G to $\sim 10^{15}$G 
during merger \cite{Rezzolla:2011da}.

\subsection{Time scale summary}

These time scale estimates indicate that the neutrino cooling
time scale can be comparable to the magnetic braking/angular momentum
loss time scales in typical HMNSs. If thermal pressure is the dominant
source of support in an HMNS against catastrophic collapse to a BH,
then the cooling time scale will determine the time interval between
the GW signals at merger and collapse. Even if thermal pressure
contributes only partially to the support of the HMNS, the remnant
will collapse faster with cooling than without.
%then a model
%that accounts for both cooling and angular momentum
%redistribution/loss 
%will lead to an earlier collapse than a model where cooling is not accounted for.
These considerations  necessitate the modeling of neutrino
cooling in simulations of NSNS mergers that form HMNSs, 
not only to predict the neutrino signature, but also to determine what
mechanism drives the remnant to its final configuration. Knowing the
results from such simulations, it may be possible to extract useful
information about the temperature of the matter, neutrino cooling mechanisms, the existence of bar modes,
and the magnetic field strength and possibly place constraints on the nuclear EOS from the GW
observations. We perform preliminary simulations to probe this issue below.

\section{Basic Equations}
\label{sec:basic_eqns}

This section introduces our notation, summarizes our methods and
numerical techniques as described in~\cite{eflstb08,elsb09,els10,APPENDIXPAPER}. 
Greek indices denote all four spacetime dimensions (0, 1, 2, and 3), and Latin
indices label spatial parts only (1, 2, and 3).

We use the 3+1 formulation of general relativity and decompose
the metric into the following form:
\beq
  ds^2 = -\alpha^2 dt^2
+ \gamma_{ij} (dx^i + \beta^i dt) (dx^j + \beta^j dt) \ .
\eeq
The fundamental variables for metric evolution are the spatial
three-metric $\gamma_{ij}$ and extrinsic curvature $K_{ij}$. We adopt
the Baumgarte-Shapiro-Shibata-Nakamura (BSSN) 
formalism~\cite{SN,BS} in which 
the evolution variables are the conformal exponent $\phi
\equiv \ln (\gamma)/12$, the conformal 3-metric $\tilde
\gamma_{ij}=e^{-4\phi}\gamma_{ij}$, three auxiliary functions
$\tilde{\Gamma}^i \equiv -\tilde \gamma^{ij}{}_{,j}$, the trace of
the extrinsic curvature $K$, and the trace-free part of the conformal extrinsic
curvature $\tilde A_{ij} \equiv e^{-4\phi}(K_{ij}-\gamma_{ij} K/3)$.
Here, $\gamma={\rm det}(\gamma_{ij})$. The full spacetime metric $g_{\mu \nu}$
is related to the three-metric $\gamma_{\mu \nu}$ by $\gamma_{\mu \nu}
= g_{\mu \nu} + n_{\mu} n_{\nu}$, where the future-directed, timelike
unit vector $n^{\mu}$ normal to the time slice can be written in terms
of the lapse $\alpha$ and shift $\beta^i$ as $n^{\mu} = \alpha^{-1}
(1,-\beta^i)$. Evolution equations for these BSSN variables are 
given by Eqs.~(9)--(13) in~\cite{eflstb08}. 
We adopt the standard puncture gauge conditions: an advective
``1+log'' slicing condition for the lapse and a
``$\Gamma$-freezing'' condition for the shift~\cite{GodGauge}. The
evolution equations for $\alpha$ and $\beta^i$ are given by
Eqs.~(2)--(4) in~\cite{elsb09}, with the
$\eta$ parameter set to $0.2/M$,
where $M$ is the ADM mass of the NSNS binary.
We add a fifth-order Kreiss-Oliger dissipation term to all evolved
BSSN, lapse and shift variables to reduce high-frequency numerical
noise associated with AMR refinement interfaces.

The fundamental hydrodynamic (HD) variables are the rest-mass density 
$\rho_0$, specific internal energy $\epsilon$, pressure $P$, and
four-velocity $u^{\mu}$.
We adopt a $\Gamma$-law equation of state (EOS)
$P=(\Gamma-1)\rho_0 \epsilon$ with $\Gamma=2$, which reduces to
an $n=1$ polytropic law $[P=\kappa \rho_0^{(1+1/n)}]$ for the initial (cold) neutron star matter.
The fluid stress-energy tensor is given by
\beq
  T_{\mu \nu} = \rho_0 h u_\mu u_\nu + P g_{\mu \nu}\ ,
\eeq
where $h=1+\epsilon+P/\rho_0$ is the specific enthalpy.

In the standard numerical implementation of the general relativistic hydrodynamic (GRHD)
equations using a conservative scheme, it is useful to introduce the 
``conservative'' variables 
$\rho_*$, $\tilde{S}_i$, $\tilde{\tau}$. They are defined as
\beqn
&&\rho_* \equiv - \sqrt{\gamma}\, \rho_0 n_{\mu} u^{\mu} \ ,
\label{eq:rhos} \\
&& \tilde{S}_i \equiv -  \sqrt{\gamma}\, T_{\mu \nu}n^{\mu} \gamma^{\nu}_{~i}
\ , \\
&& \tilde{\tau} \equiv  \sqrt{\gamma}\, T_{\mu \nu}n^{\mu} n^{\nu} - \rho_* \ .
\label{eq:S0} 
\eeqn
The evolution equations for $\rho_*$, $\tilde{S}_i$ and $\tilde{\tau}$ can 
be derived from the conservation of rest mass $\nabla_\mu (\rho_* u^\mu)=0$ 
and the conservation of energy-momentum 
$\nabla_\mu T^{\mu \nu}=0$, giving rise to 
Eqs.~(27)--(30) in~\cite{els10}. 

\section{Numerical Methods}
\label{sec:numerical}

\subsection{Initial data}
\label{sec:initialdata}

For initial data we choose an irrotational NSNS system in a
quasiequilibrium circular orbit that consists of equal-mass, $n=1$
polytropic NSs.
The initial data satisfy the conformal thin sandwich equations \cite{BSBook}, have been calculated 
using the {\tt LORENE} spectral methods numerical libraries \cite{web:Lorene} and are publicly available. 
These data apply to a configuration with arbitrary $\kappa$, compaction (in isolation) $M/R=0.12$, 
where the compaction of the maximum mass configuration is $M/R=0.216$. Each star has a 
rest mass that is $72\%$ of the maximum allowable 
TOV rest mass for this EOS. The initial cold configuration has a coordinate separation of $11.31M$, where $M$ is the 
ADM mass of system, with $M\Omega = 0.024$, where $\Omega$ is the angular frequency of the system. The 
ADM angular momentum of the system is $J/M^2= 1.02$.
We note here that our initial data correspond to case \mbox{$1.46$-$45$-$\ast$} 
of \cite{2008PhRvD..78h4033B} and that they can be considered as the polytropic counterpart 
of case H studied in \cite{2011PhRvL.107e1102S}. If we set $\kappa = 393.9\rm~km^2$,
the ADM mass of our stars in isolation becomes $1.59 M_\odot$,
which is very close to the ADM mass ($1.6M_\odot$) in isolation of
case H in \cite{2011PhRvL.107e1102S}, where a finite temperature EOS was adopted that yields 
for zero-temperature matter a maximum TOV mass of $2.2 M_\odot$.

%%%%%%%%%%%%%%%%%%%%%%%%%%%%%%%%%%%%%%%%%%%%%%%%%%%%%%%%%%%%%%%%%%%%%%%%%%%%%%%%%%%%%%%%%%%%%%%%%%
%                                           						Figure 1
%%%%%%%%%%%%%%%%%%%%%%%%%%%%%%%%%%%%%%%%%%%%%%%%%%%%%%%%%%%%%%%%%%%%%%%%%%%%%%%%%%%%%%%%%%%%%%%%%%
\begin{figure*}%[h]
\centering
\subfigure{\includegraphics[width=0.325\textwidth]{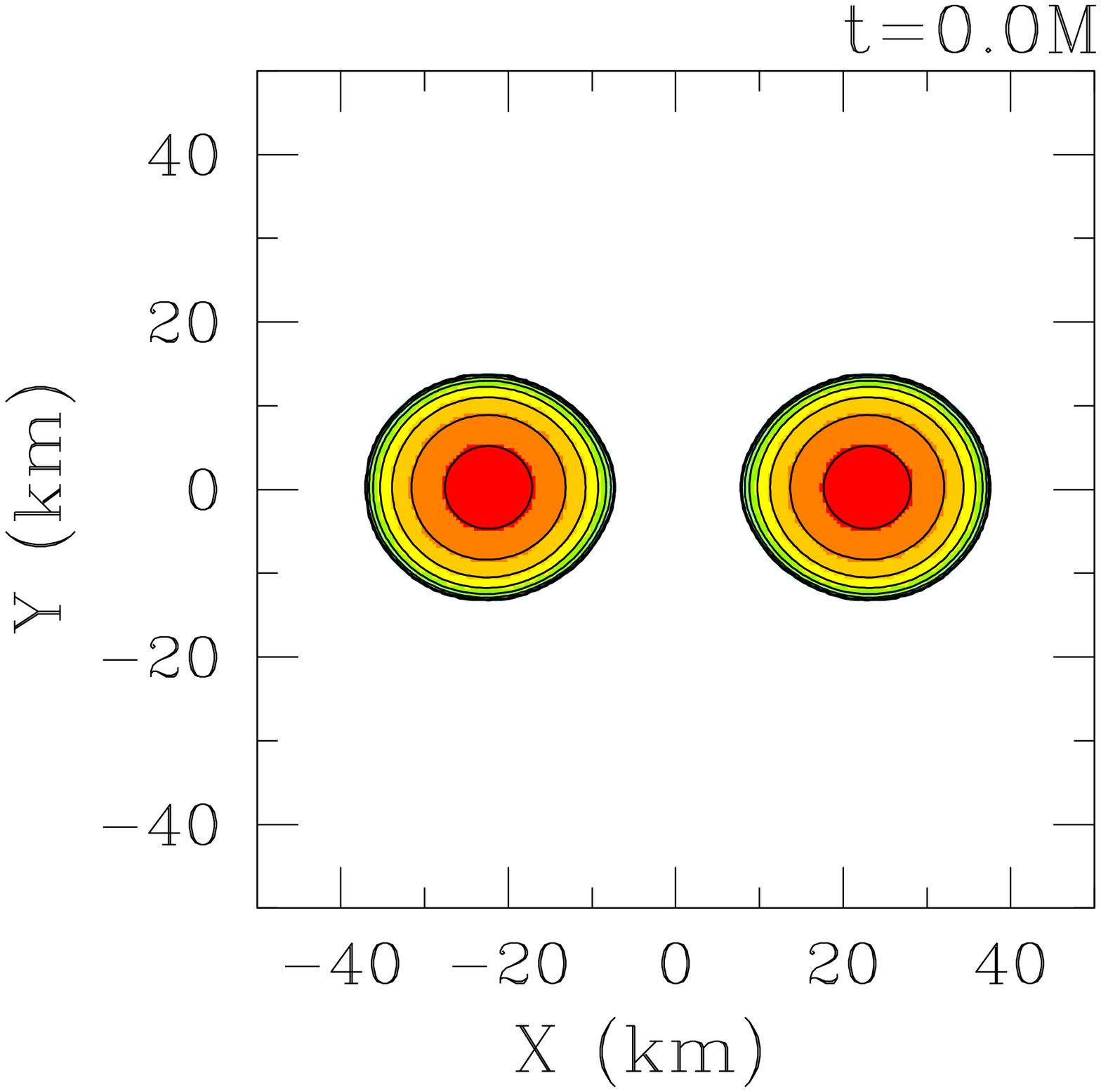}}
\subfigure{\includegraphics[width=0.325\textwidth]{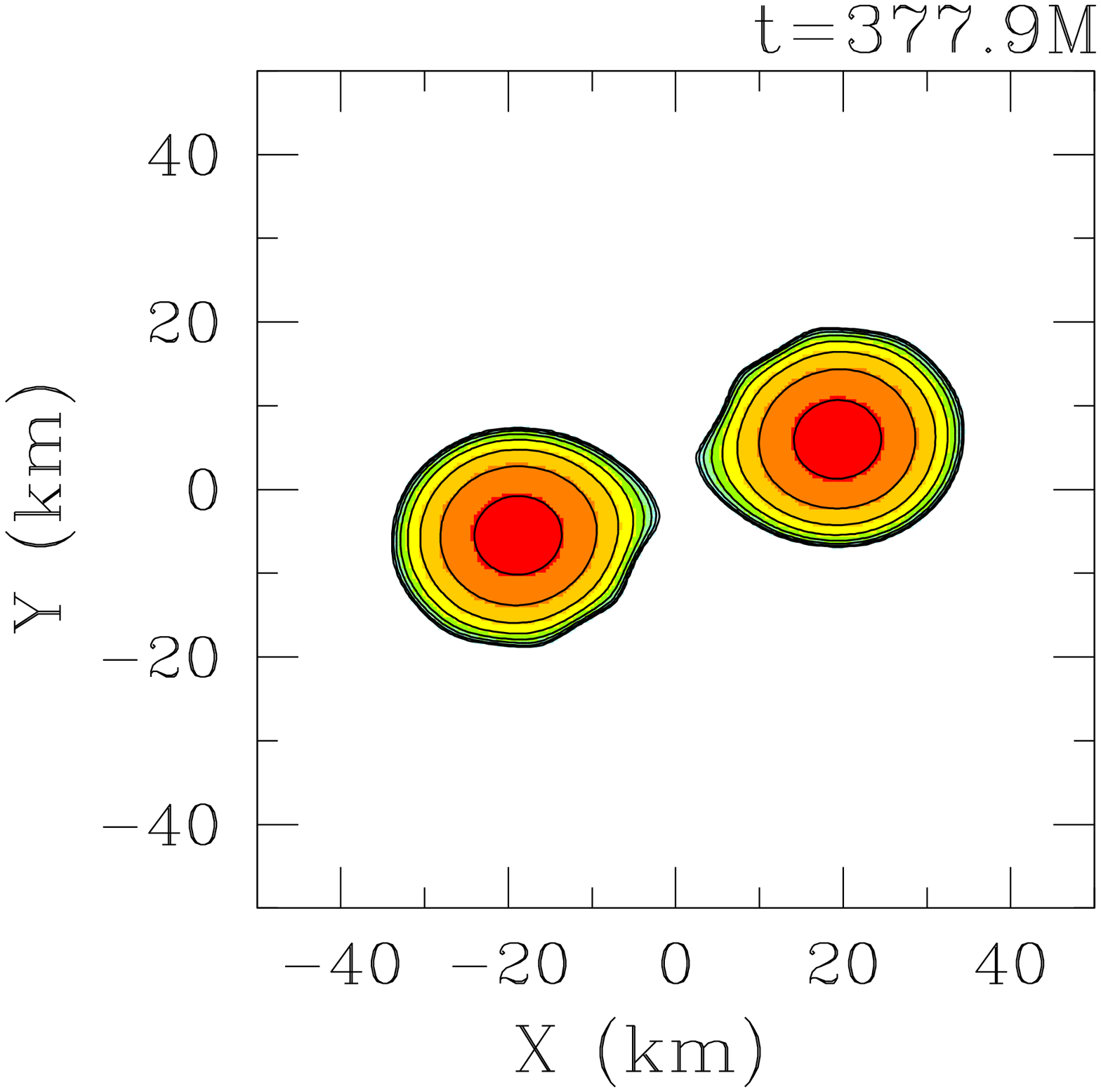}}
\subfigure{\includegraphics[width=0.325\textwidth]{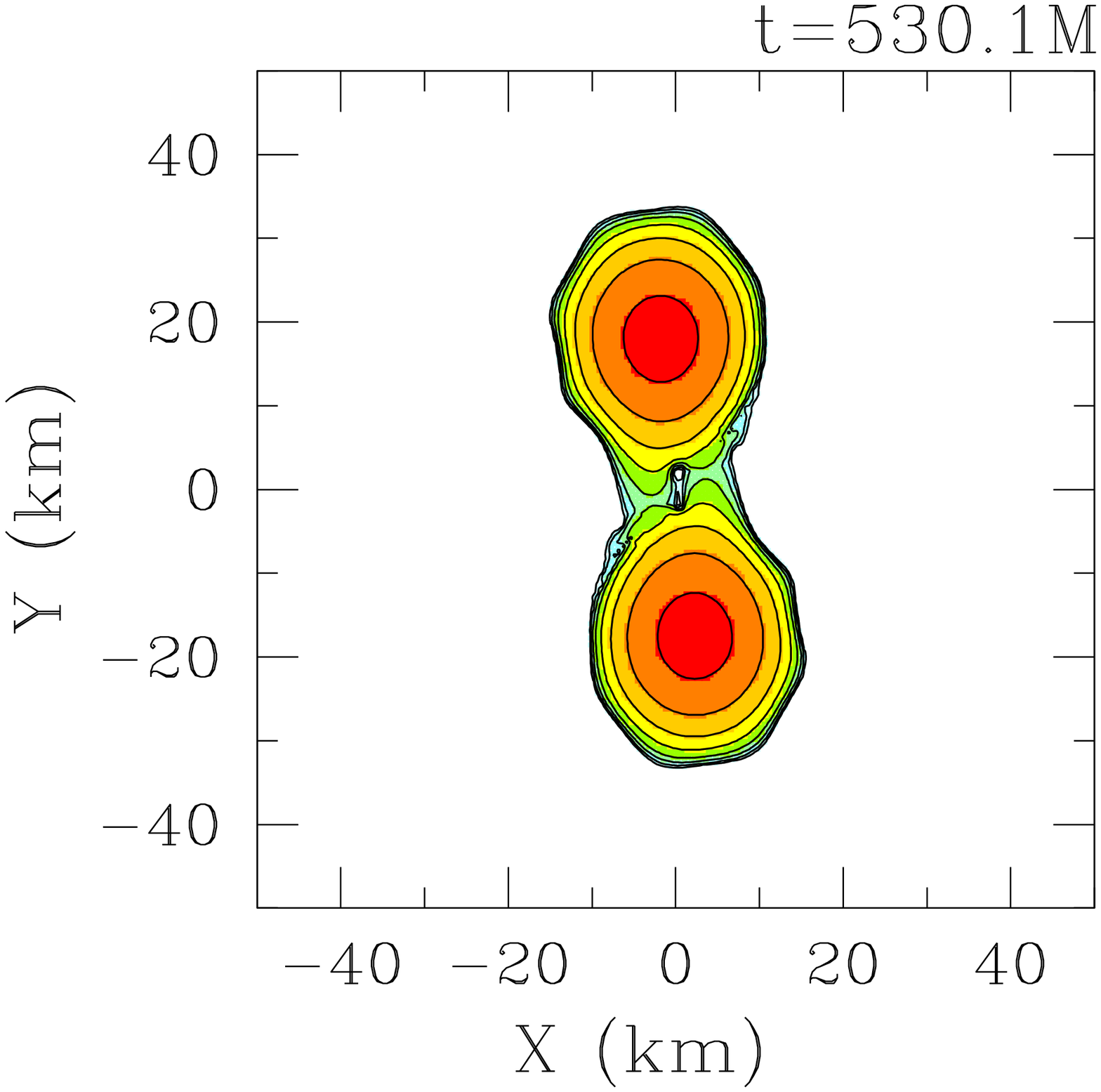}}
\subfigure{\includegraphics[width=0.325\textwidth]{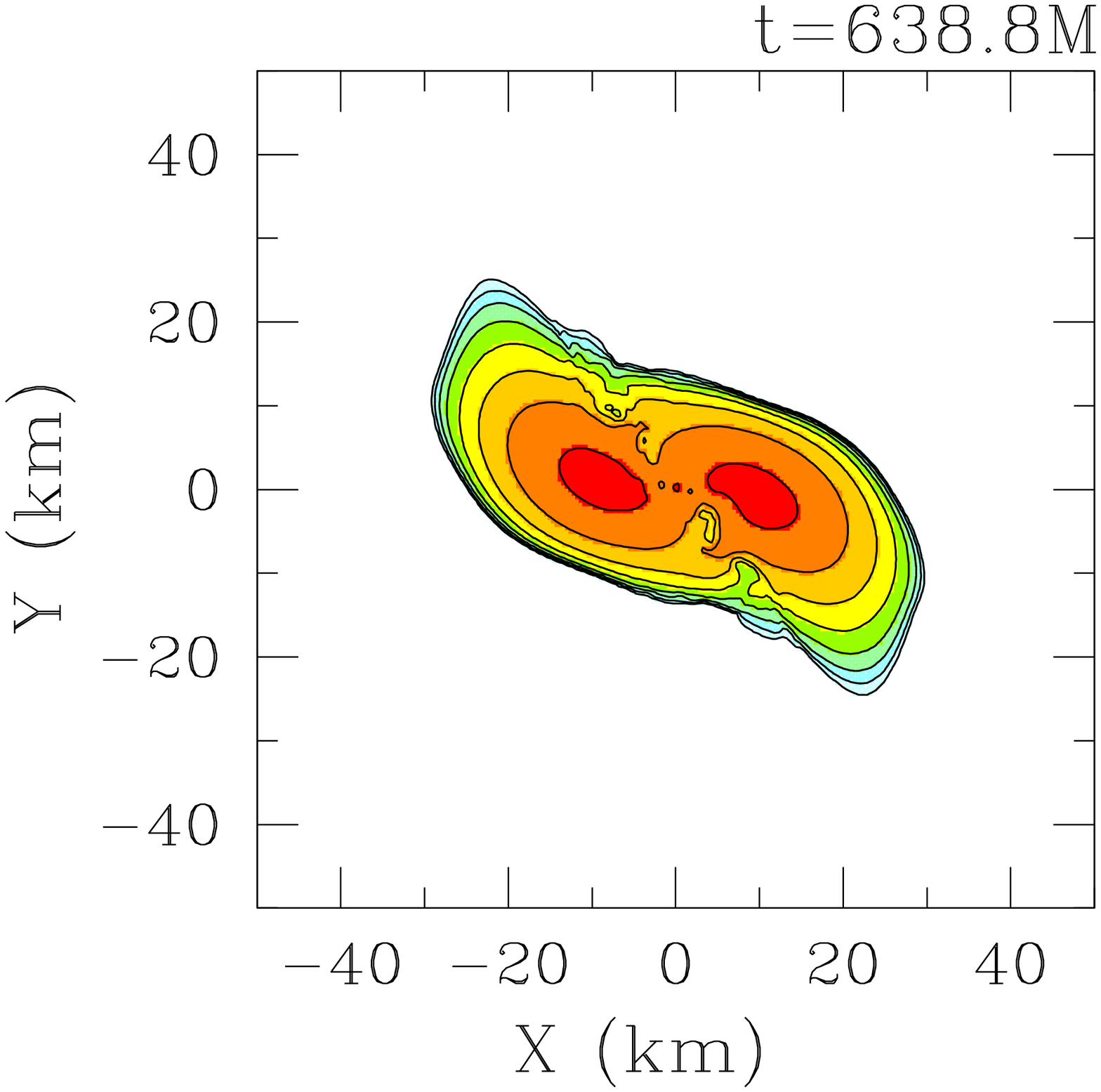}}
\subfigure{\includegraphics[width=0.325\textwidth]{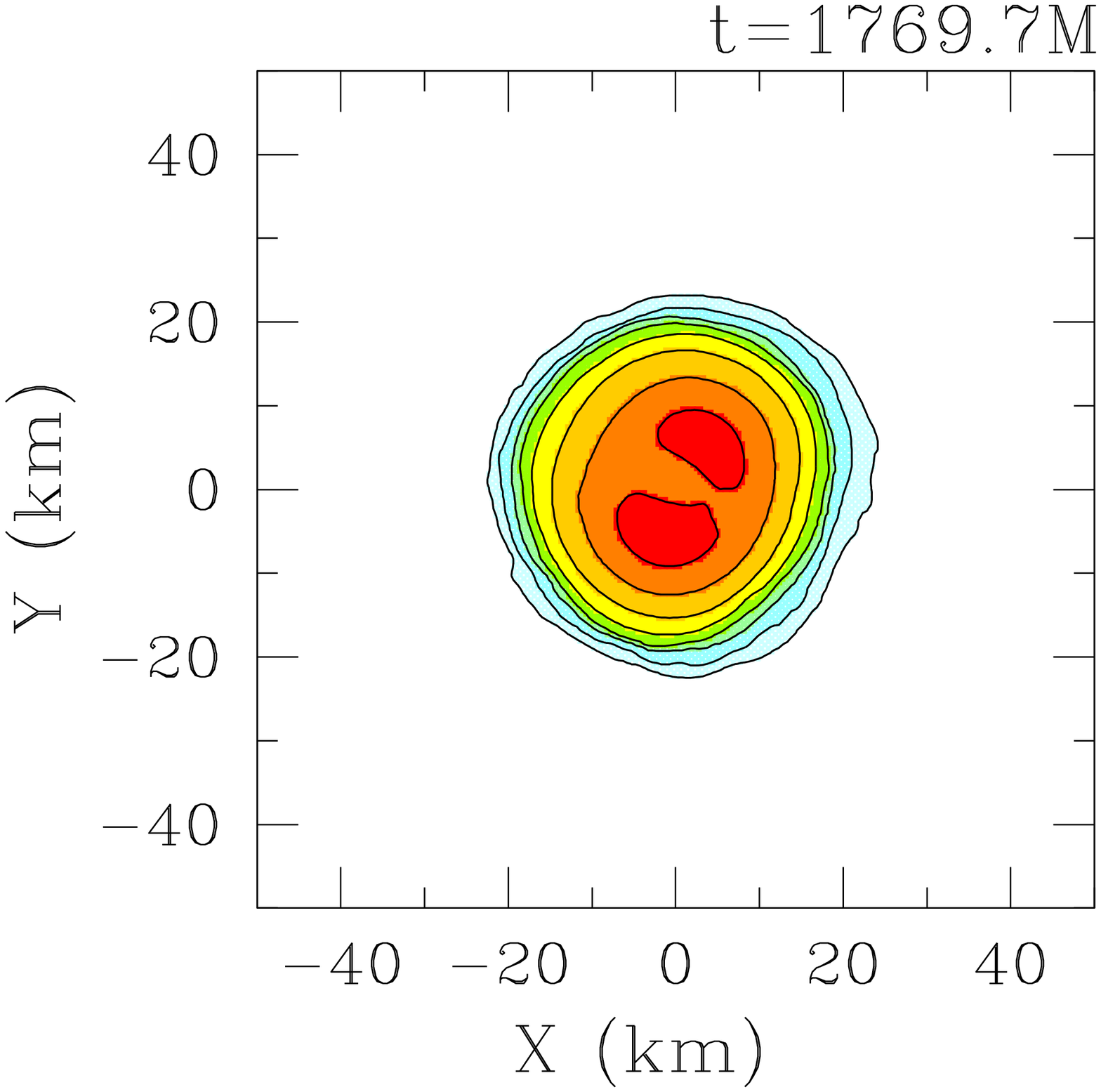}}
\subfigure{\includegraphics[width=0.325\textwidth]{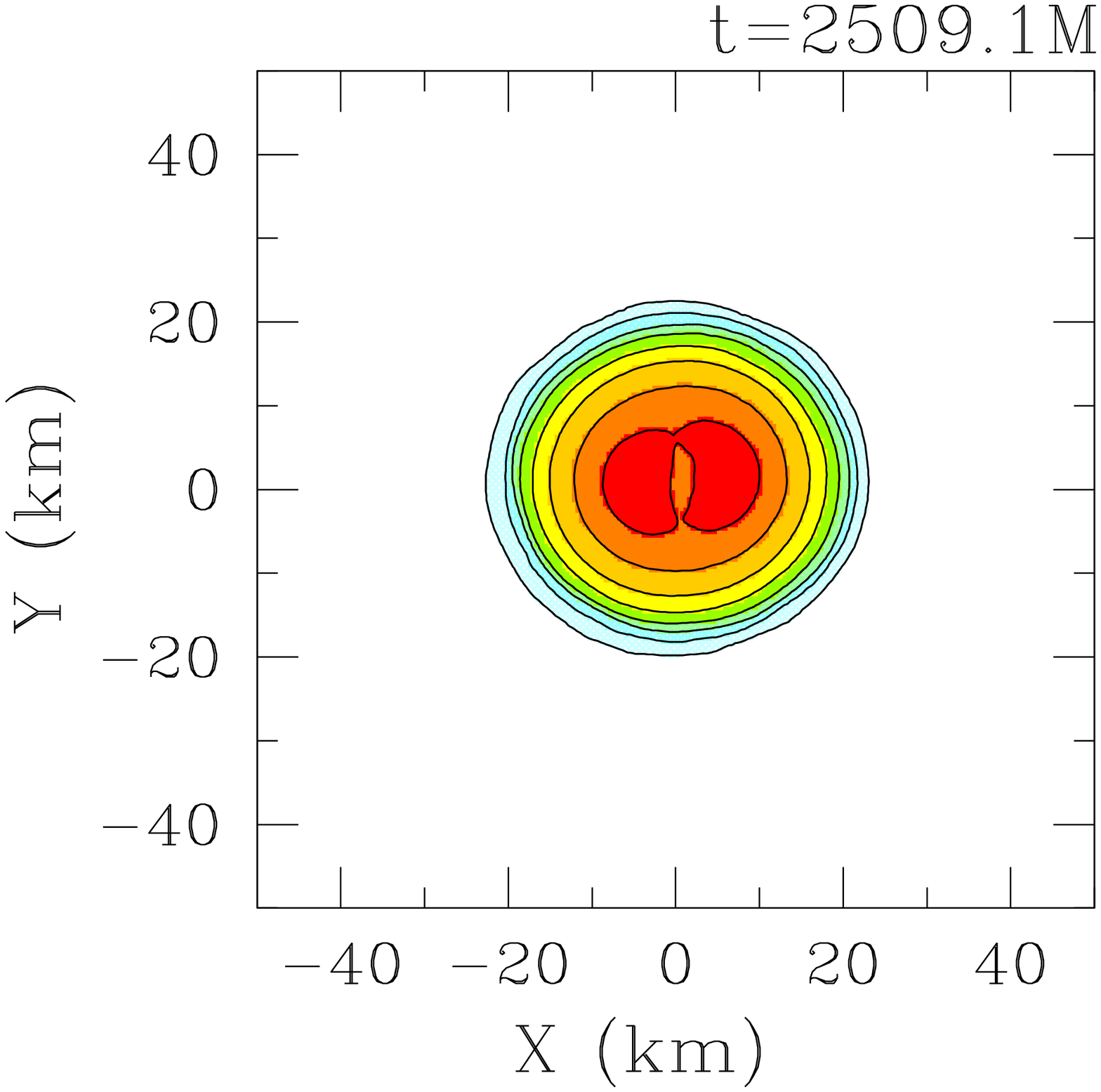}}
\caption{Case A orbital-plane rest-mass density contours at selected times. Contours are plotted according to
 $\rho_0 =  \rho_{0,{\rm max}} (10^{-0.375j-0.131})$,
  ($j$=0, 1, ... 8). The color sequence dark red, red, orange, yellow, green, light green, blue 
and light blue implies a sequence from higher to lower values.
This roughly corresponds to darker grey-scaling for higher values.  
The maximum initial NS density is
 $\kappa \rho_{0,{\rm max}} = 0.0917$, or $\rho_{0,{\rm max}}=4.58\times 10^{14}\mbox{g cm}^{-3}(1.45M_\odot/M_0)^2$. 
 Here $M=1.32\times 10^{-5} (M_0/1.45M_\odot)$s$=3.98(M_0/1.45M_\odot)$km is the ADM mass, and $M_0$ denotes
the rest mass of each star.}
\label{rho:evolution_story}
\centering
\end{figure*}
%%%%%%%%%%%%%%%%%%%%%%%%%%%%%%%%%%%%%%%%%%%%%%%%%%%%%%%%%%%%%%%%%%%%%%%%%%%%%%%%%%%%%%%%%%%%%%%%%%%
%%%%%%%%%%%%%%%%%%%%%%%%%%%%%%%%%%%%%%%%%%%%%%%%%%%%%%%%%%%%%%%%%%%%%%%%%%%%%%%%%%%%%%%%%%%%%%%%%%

\subsection{Evolution of the metric and matter}
\label{sec:num_metric_hydro}

We evolve the BSSN equations
with fourth-order accurate, centered finite-differencing stencils,
except on shift advection terms, where we use fourth-order accurate
upwind stencils.  We apply Sommerfeld outgoing wave boundary
conditions to all BSSN fields.  Our code is embedded in
the CACTUS parallelization framework~\cite{Cactus}, and our
fourth-order Runge-Kutta timestepping is managed by the {\tt MoL}
(Method of Lines) thorn, with a Courant-Friedrichs-Lewy (CFL) factor
set to 0.45. %We use the puncture gauge conditions [Eq.~(4) in~\cite{elsb09}]
%and set the parameter $\eta$ in the shift equation equal to 0.5
We use the Carpet~\cite{Carpet} infrastructure to implement the moving-box
adaptive mesh refinement. In all AMR simulations presented here, we
use second-order temporal prolongation, coupled with fifth-order
spatial prolongation. 
%The apparent horizon (AH) of the
%BH is computed with the {\tt AHFinderDirect} Cactus
%thorn~\cite{ahfinderdirect}.

The GRHD equations are evolved via a high-resolution
shock-capturing (HRSC) technique~\cite{DLSS} that employs 
PPM~\cite{PPM} coupled to
the Harten, Lax, and van Leer (HLL) approximate Riemann solver~\cite{HLL}.
The adopted GRHD scheme is second-order accurate for smooth
flows, and first-order accurate when discontinuities (e.g.\ shocks)
arise. To stabilize our scheme in regions where there is no
matter, we maintain a tenuous atmosphere on our grid, with a density
floor $\rho_{\rm atm}$ set equal to $10^{-10}$ times the initial
maximum density on our grid. The initial atmospheric pressure
$P_{\rm atm}$ is set equal to the cold
polytropic value $P_{\rm atm} = \kappa \rho_{\rm atm}^{\Gamma}$.
Throughout the
evolution, we impose limits on the atmospheric pressure to prevent
spurious heating and negative values of the internal energy
$\epsilon$ due to numerical errors. Specifically, we require $P_{\rm min}\leq P \leq P_{\rm max}$, 
where $P_{\rm max}=10 \kappa \rho_0^\Gamma$ and $P_{\rm min}=\kappa 
\rho_0^\Gamma/2$. Whenever $P$ exceeds $P_{\rm max}$ or drops below $P_{\rm min}$, we 
reset $P$ to $P_{\rm max}$ or $P_{\rm min}$, respectively.  
We impose these pressure limits only in regions where the 
rest-mass density remains very low ($\rho_0 < 100\rho_{\rm atm}$),
as in~\cite{elsb09}.

%%%%%%%%%%%%%%%%%%%%%%%%%%%%%%%%%%%%%%%%%%%%%%%%%%%%%%%%%%%%%%%%%%%%%%%%%%%%%%%%%%%%%%%%%%%%%%%%%%
%                                           						Figure 2
%%%%%%%%%%%%%%%%%%%%%%%%%%%%%%%%%%%%%%%%%%%%%%%%%%%%%%%%%%%%%%%%%%%%%%%%%%%%%%%%%%%%%%%%%%%%%%%%%%
\begin{figure}%[h]
\centering
\subfigure{\includegraphics[width=0.445\textwidth]{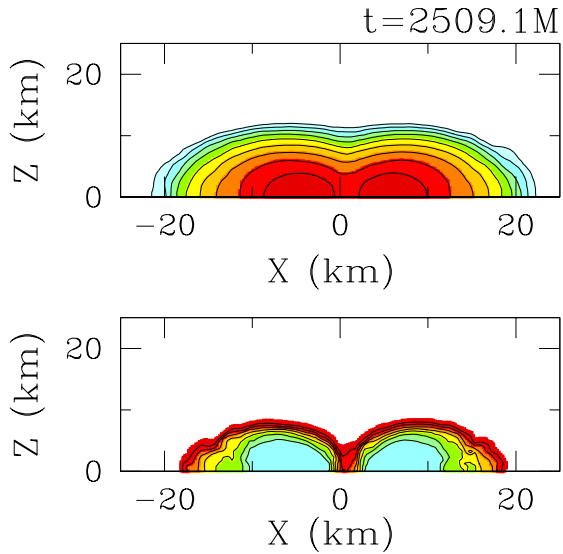}}
\caption{Case A meridional ($XZ$) plane rest-mass density (upper panel) and $K$ contours (lower panel). 
Density contours are plotted according to $\rho_0 =  \rho_{0,{\rm max}} (10^{-0.375j-0.131})$,
($j$=0, 1, ... 8), where $\kappa \rho_{0,{\rm max}} = 0.0917$, 
or $\rho_{0,{\rm max}}=4.58\times 10^{14}\mbox{g cm}^{-3}(1.45M_\odot/M_0)^2$.
 $K$ contours are plotted according to $K = K_{\rm max}10^{-0.031 j}$,
($j$=0, 1, ... 8). Here $K_{\rm max} = 1.77$. The color coding is the same as used 
in Fig.~\ref{rho:evolution_story}. In the lower panel light blue indicates 
$K\approx 1$ and dark red $K\approx1.6$. 
Here $M=1.32\times 10^{-5} (M_0/1.45M_\odot)$s$=3.98(M_0/1.45M_\odot)$km is the ADM mass, and $M_0$ denotes
the rest mass of each star.}
\label{HMNSxz}
\centering
\end{figure}
%%%%%%%%%%%%%%%%%%%%%%%%%%%%%%%%%%%%%%%%%%%%%%%%%%%%%%%%%%%%%%%%%%%%%%%%%%%%%%%%%%%%%%%%%%%%%%%%%%
%%%%%%%%%%%%%%%%%%%%%%%%%%%%%%%%%%%%%%%%%%%%%%%%%%%%%%%%%%%%%%%%%%%%%%%%%%%%%%%%%%%%%%%%%%%%%%%%%%

\subsection{Radiative cooling}

We now briefly describe our method for implementing cooling 
in our simulations. For a derivation and details regarding this
covariant cooling method, see~\cite{PLES2011}.

The dynamics of radiation is governed by \cite{MihalasBook,CollapseShapiro1996,BFarris2008}
\labeq{rad_dyn}{
\nabla_\alpha R^{\alpha\beta}=-G^{\beta},
}
where $R^{\alpha\beta}$ is the radiation stress-energy tensor given by
\labeq{}{
R^{\alpha\beta}=\int d\nu d\Omega I_{\nu}N^{\alpha}N^{\beta},
}
and $G^{\alpha}$ is the radiation four-force density given by
\labeq{}{
G^{\alpha}=\int d\nu d\Omega (\chi_\nu I_{\nu}-j_\nu)N^{\alpha}.
}
In the equations above $d\Omega$ is the solid angle, $\nu$ and $I_{\nu}=I_{\nu}(x^\alpha,N^i,\nu)$ 
are the radiation frequency and
specific intensity of radiation at $x^\alpha$ moving in direction $N^\alpha=p^\alpha/h\nu$, respectively. 
All quantities are measured in the local Lorentz frame of a 
fiducial observer with four-velocity
$u^{\alpha}_{fid}$, i.e., 
\labeq{}{
h\nu = -p_{\alpha} u^{\alpha}_{fid},
}
where $p^{\alpha}$ is the photon four-momentum and $h$ denotes Planck's constant. The
energy-momentum conservation equation then becomes
\labeq{}{
\nabla_{\alpha} (T^{\alpha\beta}+R^{\alpha\beta}) =0
}
or after using Eq.~\eqref{rad_dyn}
\labeq{divT}{
\nabla_{\alpha} T^{\alpha\beta} =G^{\beta}.
}

Our artificial cooling prescription amounts to finding 
a functional form for $G^{\beta}$ such that thermal energy and pressure
are drained from the system. Choosing
\labeq{isoG}{
G^{\alpha}=-u^{\alpha}\Lambda,
}
and setting
\labeq{emis}{
\Lambda = \frac{\rho_0}{\tau_c}\epsilon_{\rm th},
}
where $\tau_c$ is some prescribed cooling time scale, it can be shown
that in a frame comoving with the fluid the specific thermal energy of a fluid
parcel evolves as follows \cite{PLES2011}
\labeq{thermal2a}{
\frac{d}{d\tau}\epsilon_{\rm th} = \bigg[\frac{(\Gamma_{\rm th}-1)}{\rho_0}\frac{d\rho_0}{d\tau}-\frac{1}{\tau_c}\bigg]\epsilon_{\rm th},
}
where $\tau$ is the proper time of a comoving observer.

The first term in brackets on the RHS of Eq.~\eqref{thermal2a} arises from 
adiabatic compression or expansion. 
The second term corresponds to cooling and radiates away thermal energy exponentially. 

Projecting Eq.~\eqref{divT} using the timelike unit vector $n^{\alpha}$ 
normal to spacelike hypersurfaces and the projection operator 
$h^{\alpha}{}_{\beta}=\delta^{\alpha}{}_{\beta} + n^{\alpha} n_{\beta}$, we find that
the 3+1 GRHD equations become
\labeq{Stilde}{
\partial_t \tilde S_i +\partial_j(\alpha\sqrt{\gamma}T^j{}_{i})=
               \frac{1}{2}\alpha\sqrt{\gamma} T^{\alpha\beta}g_{\alpha\beta,i}
	       -\alpha\sqrt{\gamma}u_i \Lambda,
}
and 
\labeq{tautilde}{
\partial_t \tilde \tau +\partial_i(\alpha^2\sqrt{\gamma}T^{0i}-\rho_*v^i)= s
	       -\alpha^2\sqrt{\gamma}u^0 \Lambda,
}
where we have used Eq.~\eqref{isoG}. Thus 
cooling enters as a source term in the GRHD equations.

\subsection{Recovery of primitive variables}
\label{sec:inversion}

At each timestep,
we need to recover the ``primitive variables''
$\rho_0$, $P$, and $v^i$ from the ``conservative'' variables
$\rho_*$, $\tilde{\tau}$, and $\tilde{S}_i$.
We perform the inversion by numerically solving two nonlinear
equations via the Newton-Raphson method as described in~\cite{ngmz06},
using the code developed in~\cite{harmsolver}.

Sometimes the ``conservative'' variables may assume values which are 
out of physical range, resulting in unphysical primitive
variables after inversion (e.g.\ negative pressure or even
complex solutions). This usually happens in the low-density 
``atmosphere'' or deep inside the BH interior (when a BH is present) 
where high-accuracy evolution is difficult to maintain. Various techniques have been 
suggested to handle the inversion failure (see, e.g.~\cite{bs2011}). 
Our approach is mainly to impose constraints on the conservative 
variables to reduce the inversion failure. For a summary of our latest 
techniques, see \cite{Etienne:2011ea}.

\subsection{Diagnostics}
\label{sec:diagnostics}

\subsubsection{Constraints and rest-mass conservation}
During the evolution, we monitor the Hamiltonian and momentum
constraints, calculated by Eqs.~(40)--(43) of~\cite{eflstb08}.

%%%%%%%%%%%%%%%%%%%%%%%%%%%%%%%%%%%%%%%%%%%%%%%%%%%%%%%%%%%%%%%%%%%%%%%%%%%%%%%%%%%%%%%%%%%%%%%%%%
%                                           						Figure 3
%%%%%%%%%%%%%%%%%%%%%%%%%%%%%%%%%%%%%%%%%%%%%%%%%%%%%%%%%%%%%%%%%%%%%%%%%%%%%%%%%%%%%%%%%%%%%%%%%%
\begin{figure*}%[h]
\centering
\subfigure{\includegraphics[width=0.325\textwidth]{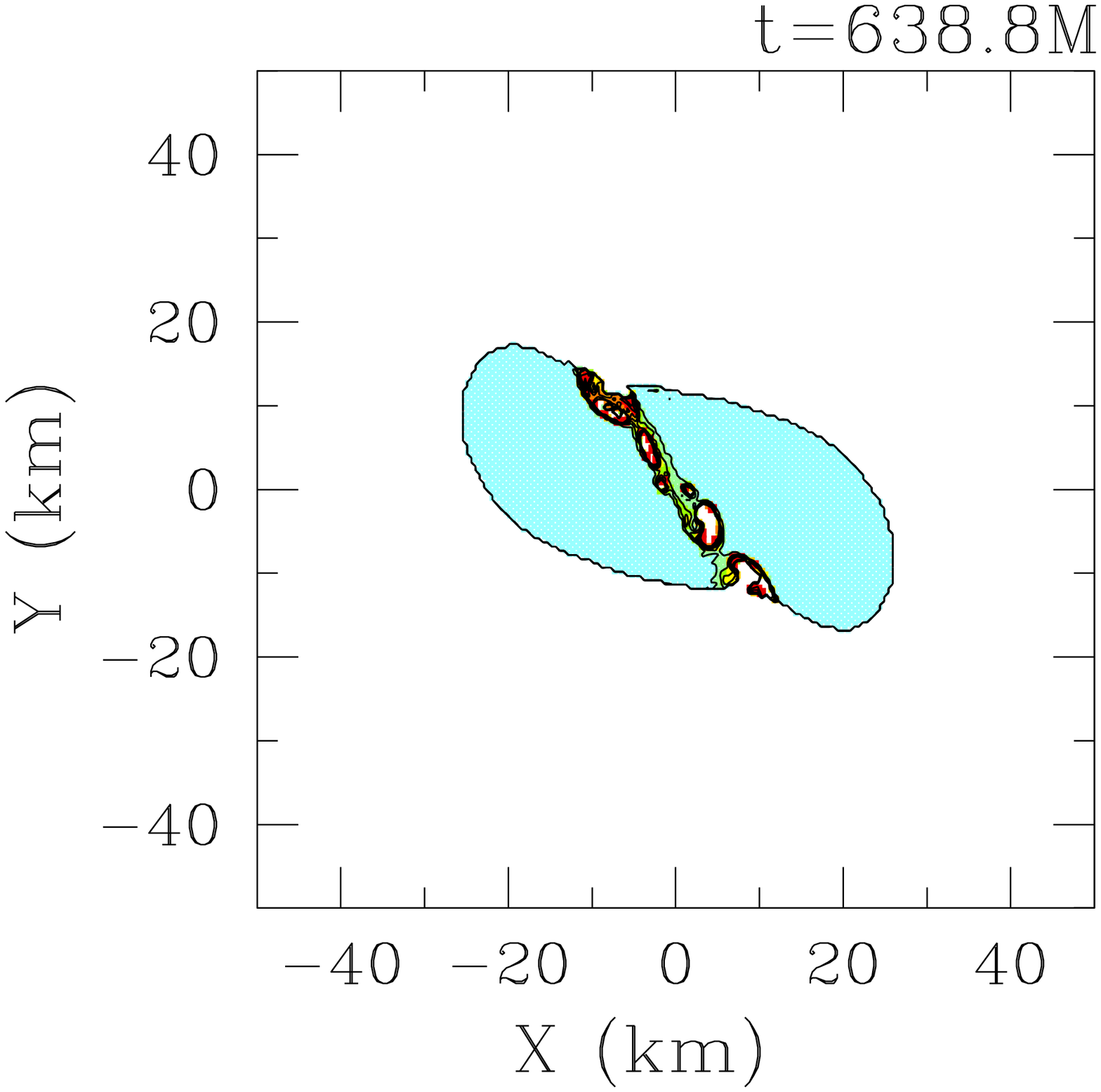}}
\subfigure{\includegraphics[width=0.325\textwidth]{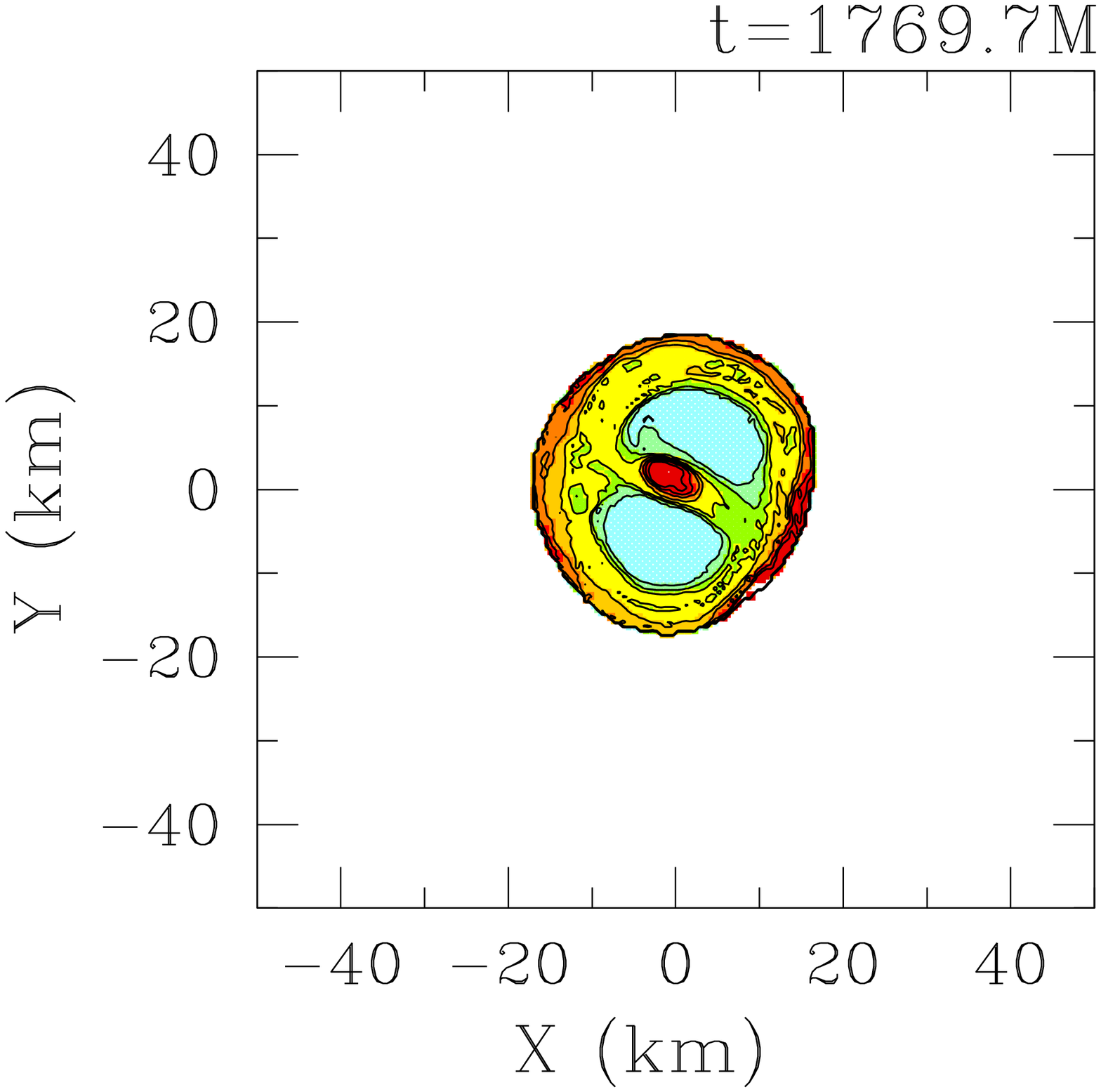}}
\subfigure{\includegraphics[width=0.325\textwidth]{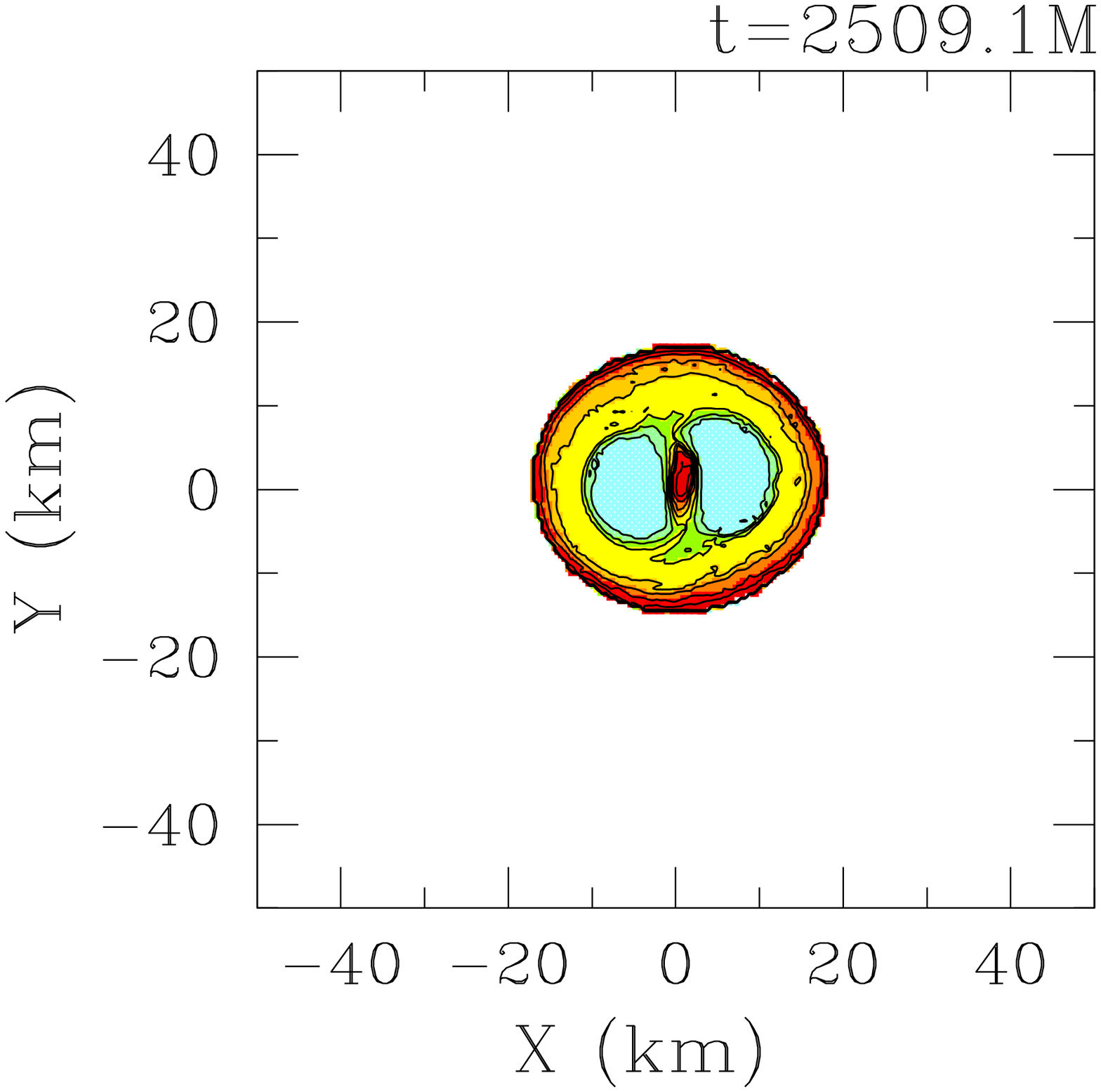}}
\caption{Case A orbital-plane $K$ contours at selected times. Contours are plotted according to $K = K_{\rm max}10^{-0.028 j}$,
  ($j$=0, 1, ... 8). Here $K_{\rm max} = 1.6$. The color coding is the same as 
used in Figs.~\ref{rho:evolution_story} and~\ref{HMNSxz}.
A density cutoff of $10^{-1}\rho_{0,max}$ has been imposed, where
 $\kappa \rho_{0,{\rm max}} = 0.0917$, or $\rho_{0,{\rm max}}=4.58\times 10^{14}\mbox{g cm}^{-3}(1.45M_\odot/M_0)^2$.
 The dual cold core nature of the HMNS is visible and it becomes clear that between the two cores a hot area has formed, 
where 40-50\% of the total pressure is due to thermal pressure. In the outer parts of the HMNS the contribution
of the thermal component is greater than 50\% of the total pressure.
 Here $M=1.32\times 10^{-5} (M_0/1.45M_\odot)$s$=3.98(M_0/1.45M_\odot)$km is the ADM mass, and $M_0$ denotes
the rest mass of each star.}
\label{K:evolution_story}
\centering
\end{figure*}
%%%%%%%%%%%%%%%%%%%%%%%%%%%%%%%%%%%%%%%%%%%%%%%%%%%%%%%%%%%%%%%%%%%%%%%%%%%%%%%%%%%%%%%%%%%%%%%%%%
%%%%%%%%%%%%%%%%%%%%%%%%%%%%%%%%%%%%%%%%%%%%%%%%%%%%%%%%%%%%%%%%%%%%%%%%%%%%%%%%%%%%%%%%%%%%%%%%%%

%%%%%%%%%%%%%%%%%%%%%%%%%%%%%%%%%%%%%%%%%%%%%%%%%%%%%%%%%%%%%%%%%%%%%%%%%%%%%%%%%%%%%%%%%%%%%%%%%%
%                                           						Figure 5
%%%%%%%%%%%%%%%%%%%%%%%%%%%%%%%%%%%%%%%%%%%%%%%%%%%%%%%%%%%%%%%%%%%%%%%%%%%%%%%%%%%%%%%%%%%%%%%%%%
\begin{figure*}[t]
\centering
\subfigure{\includegraphics[width=0.325\textwidth,angle=0]{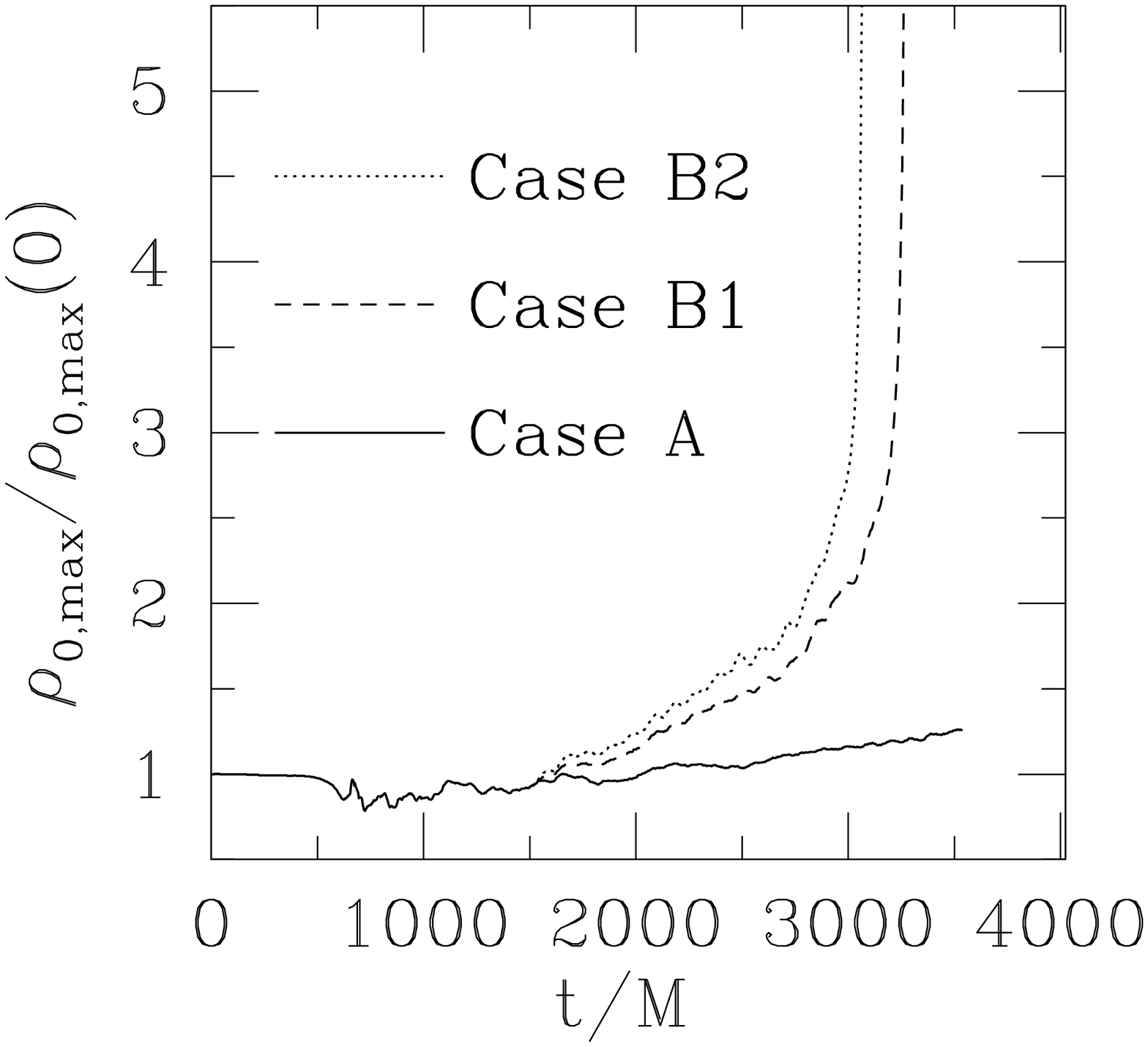}}
\subfigure{\includegraphics[width=0.325\textwidth,angle=0]{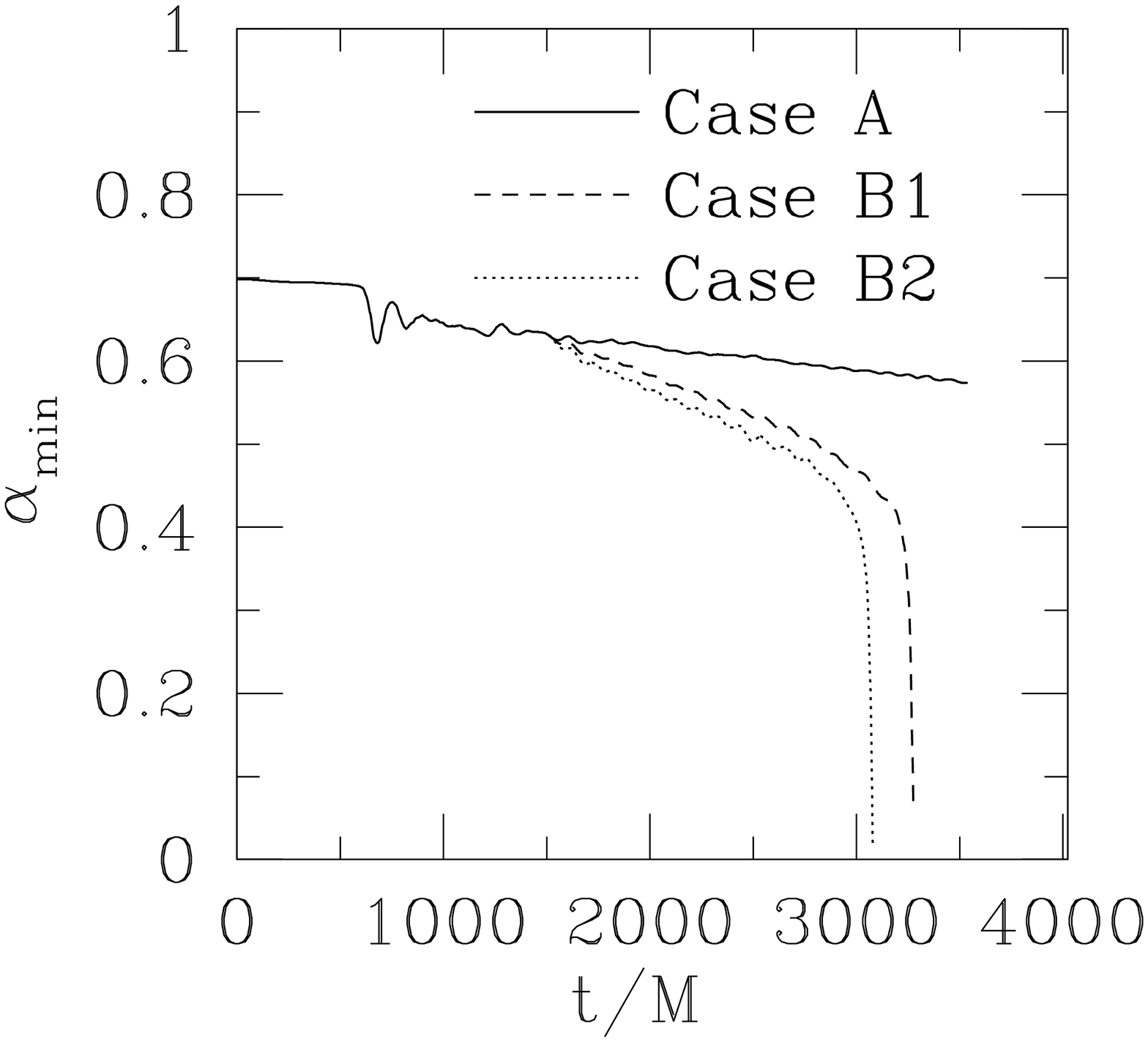}}
\subfigure{\includegraphics[width=0.325\textwidth,angle=0]{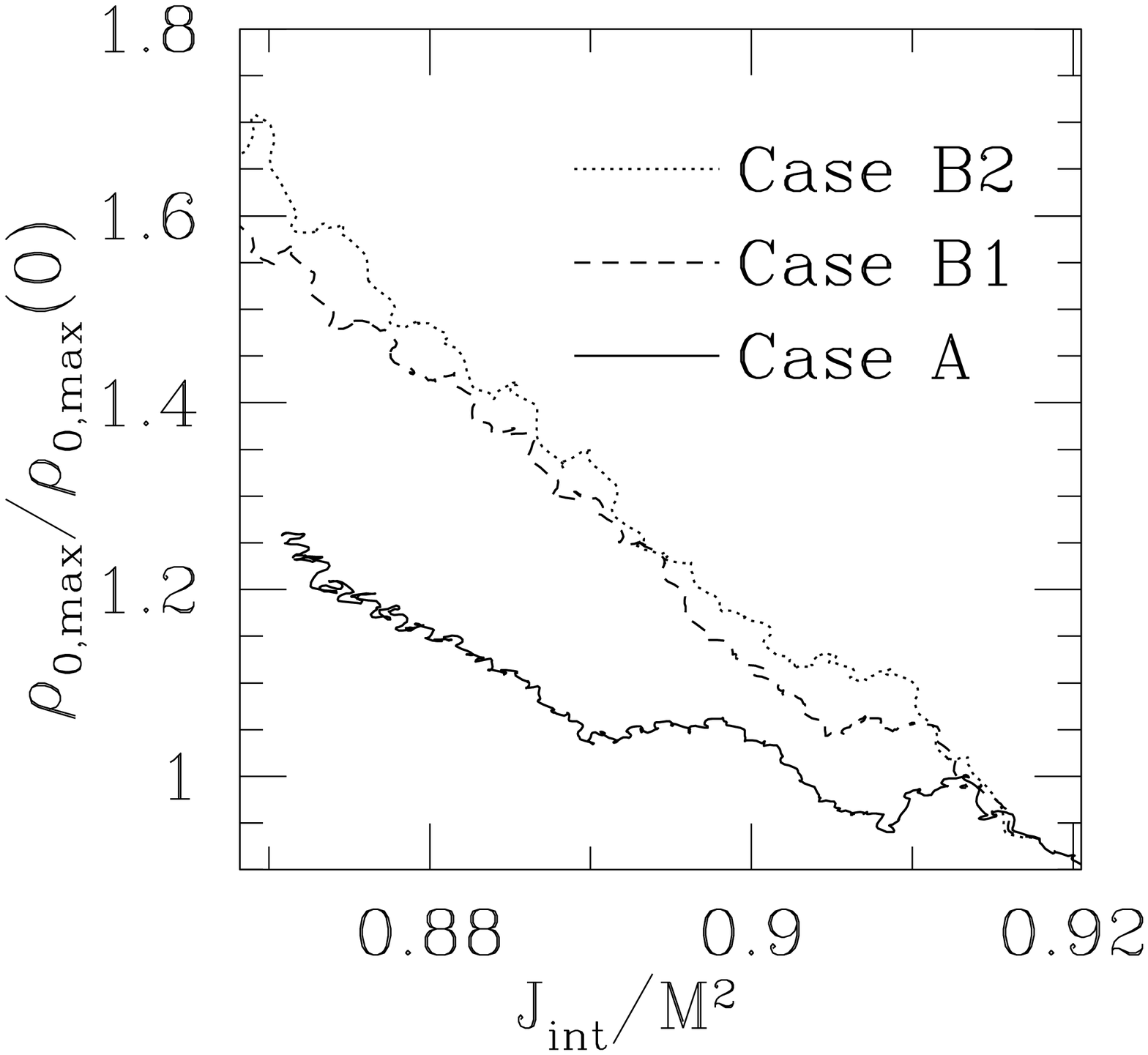}}
\caption{Left panel: Maximum rest-mass density $\rho_{0,\rm max}(t)$
  normalized by $\rho_{0,\rm max} (t=0)$. Middle panel: Minimum value of the lapse function vs time.
Right panel: Maximum rest-mass density vs total angular momentum for the different cooling time scales considered.
Here $M=1.32\times 10^{-5} (M_0/1.45M_\odot)$s$=3.98(M_0/1.45M_\odot)$km is the ADM mass, and 
 $\kappa \rho_{0,{\rm max}}(0) = 0.0917$, or $\rho_{0,{\rm max}}(0)=4.58\times 10^{14}\mbox{g cm}^{-3}(1.45M_\odot/M_0)^2$.
$M_0$ here denotes the rest mass of each star.
\label{fig:ang_vs_rhoc}
}
\centering
\end{figure*}
%%%%%%%%%%%%%%%%%%%%%%%%%%%%%%%%%%%%%%%%%%%%%%%%%%%%%%%%%%%%%%%%%%%%%%%%%%%%%%%%%%%%%%%%%%%%%%%%%%
%%%%%%%%%%%%%%%%%%%%%%%%%%%%%%%%%%%%%%%%%%%%%%%%%%%%%%%%%%%%%%%%%%%%%%%%%%%%%%%%%%%%%%%%%%%%%%%%%%

When hydrodynamic matter is evolved on a fixed uniform grid, our
hydrodynamic scheme guarantees that the rest mass $M_0$ is conserved
to machine roundoff error.  This strict conservation is no longer maintained
in an AMR grid, where spatial and temporal prolongation is performed
at the refinement boundaries.  Hence, we also monitor the total
rest mass,
\beq
  M_0 = \int \rho_* d^3x,
\label{eq:m0}
\eeq
during the evolution. Rest-mass conservation is also violated whenever 
$\rho_0$ spuriously drops below and is then reset to the atmosphere
value. This usually happens only in the very low-density atmosphere or
deep inside the BH horizon where high accuracy
is difficult to maintain. In the simulations presented in this paper 
the violation of rest-mass conservation is less than $1\%$. 

\subsubsection{Temperature}

We measure the thermal energy generated by shocks via the entropy 
parameter $K\equiv P/P_{\rm cold}$, where $P_{\rm cold}=\kappa \rho_0^\Gamma$ is
the pressure associated with the cold EOS. 
The specific internal energy can be decomposed into a
cold part and a thermal part: $\epsilon = \epsilon_{\rm cold} +
\epsilon_{\rm th}$ with 
\begin{equation}
\epsilon_{\rm cold} = -\int P_{\rm cold}
d(1/\rho_0) = \frac{\kappa}{\Gamma - 1} \rho_0^{\Gamma-1}\ .
\end{equation}
Hence the relationship between $K$ and $\epsilon_{\rm th}$ is 
\beqn
\epsilon_{\rm th} & = & \epsilon - \epsilon_{\rm cold} =
\frac{1}{\Gamma - 1} \frac{P}{\rho_0} - \frac{\kappa}{\Gamma - 1}
\rho_0^{\Gamma-1} \nonumber \\
& = & (K - 1) \epsilon_{\rm cold} \ .
\eeqn
For shock-heated gas, we always have $K>1$ (see 
Appendix~B of~\cite{elsb09}. 

To estimate the temperature of the remnant we model the specific 
thermal energy as
\labeq{temp_dep}{
\epsilon_{\rm th} = \frac{3k\rm T}{2m_{\rm n}}+ f\frac{a\rm T^4}{\rho_0},
}
where $m_{\rm n}$ is the mass of a nucleon, $k$ is Boltzmann's constant 
and $a$ is the radiation constant.
The first term represents the approximate thermal energy of the nucleons, 
and the second term accounts for the thermal energy due to 
relativistic particles. 
The factor $f$ reflects the number of species of relativistic particles that 
contribute to the thermal energy and is determined self-consistently as
outlined in \cite{PLES2011}.

Note that the value of $T$ depends on the adopted mass of the initial configuration
and breaks the scale invariance with respect to $\kappa$. For this purpose we set 
$\kappa=269.6\rm km^2$, for which $M=2.69M_\odot$.

%##################################################
%%%%%%############################ Table 1
%##################################################
\begin{table}
\caption{Summary of cases. The second column 
indicates whether cooling is applied. 
%A $\checkmark$ means ``Yes'' and a \ding{55} means ``No''.
Here $M=1.32\times 10^{-5} (M_0/1.45M_\odot)$s$=3.98(M_0/1.45M_\odot)$km is the ADM mass.
}
\begin{tabular}{ccc}
  \hline
\ Case Name$^{(a)}$\ &\ Cooling On\  &\ Cooling time scale, $\tau/M$\ \\
  \hline 
 A & No & $\infty$ \\
 B1 & Yes & $150.82$ \\
 B2 & Yes & $301.64$ \\
\hline
\end{tabular}
\begin{flushleft}
$^{(a)}$ The inspiral and merger calculation is part of case A.
In cases B1 and B2 cooling is turned on at $t\approx 1600 M$,
at which point the HMNS remnant of case A has relaxed to a quasiequilibrium 
state.
\end{flushleft}
\label{cases}
\end{table}
%##################################################

\subsubsection{GW extraction, energy and angular momentum conservation}

Gravitational waves are extracted using the Newman-Penrose Weyl scalar $\psi_4$ at
various extraction radii between $40M$ and $150M$. We decompose
$\psi_4$ into $s=-2$ spin-weighted spherical harmonics up to and including
$l=4$ modes. At each extraction radius, the retarded time is computed
using the technique described in Sec.~IIB of~\cite{bm09} to reduce the near-field effect. 

We compute the radiated energy $\Delta E_{GW}$ and $z$-component of
angular momentum $\Delta J_{GW}$ %and linear momentum $\Delta P^i_{GW}$
using expressions equivalent to Eqs.~(33), (39), (40) and (49) of~\cite{rant08}.

We also monitor the mass $M_{\rm int}$ and ($z$-component of) 
the total angular momentum $J_{\rm int}$ interior to the simulation
domain. These quantities are defined as integrals over the surface of
the outer boundary $\partial V)$ of the computational domain:
\beqn
  M_{\rm int} &=&\frac{1}{2\pi}\oint_{\partial V} \left(\frac{1}{8}\tilde{\Gamma}^i -
\tilde{\gamma}^{ij}\partial_j\psi \right)d\Sigma_i,\label{eq:mintsurf}\\
J_{\rm int}&=&\frac{1}{8\pi}{\tilde{\epsilon}_{zj}}^k\oint_{\partial V} x^j(K^m_k-\delta^m_k
K)d\Sigma_m,\label{eq:jintsurf}
\eeqn 
where $\tilde{\epsilon}_{ijk}$ is the flat-space Levi-Civita tensor.
As pointed out in~\cite{elsb09}, the integrals can be evaluated more accurately 
by alternative expressions that use Gauss's law~\cite{BSBook}: 
\beqn
  M_{\rm int}&=& \int_V d^3x \left(\psi^5\rho + {1\over16\pi}\psi^5 \tilde{A}_{ij}
\tilde{A}^{ij} - {1\over16\pi}\tilde{\Gamma}^{ijk}\tilde{\Gamma}_{jik} \right.
\ \  \cr
&& \left. + {1-\psi\over16\pi}\tilde{R} - {1\over24\pi}\psi^5K^2\right) \cr
 && + {1\over 2\pi} \oint_S
\left( \frac{1}{8}\tilde{\Gamma}^i-\tilde{\gamma}^{ij} \partial_j \psi
\right) d\Sigma_i \ , \label{eq:Mint_sur_vol}
\eeqn
\beqn
  J_{\rm int} &=& {1\over8\pi} \tilde{\epsilon}_{zj}{}^n\int_V d^3x
           \psi^6(\tilde{A}^j{}_n + {2\over3}x^j\partial_nK \cr
         && - {1\over2} x^j\tilde{A}_{km}\partial_n\tilde{\gamma}^{km}
         + 8\pi x^j \tilde{S}_n) \cr
           && + {1\over8\pi} \tilde{\epsilon}_{zj}{}^n \oint_S
           \psi^6 x^j \tilde{A}^m{}_n d\Sigma_m \ , \label{eq:Jint_sur_vol}
\eeqn
where $S$ is a surface surrounding the BH horizon (when a BH is present), 
$V$ is the volume between 
$S$ and the outer boundary, $\rho=n_\mu n_\nu T^{\mu \nu}$, and $\tilde{R}$ is 
the Ricci scalar associated with the conformal 3-metric $\tilde{\gamma}_{ij}$.

To check the violation of energy and angular momentum conservation, 
we monitor the quantities
\beqn
  \delta E &=& |M-M_{\rm int}(t)-\Delta E_{\rm GW}(t)|/M \ , \label{eq:deltae} \\
  \delta J &=& |J-J_{\rm int}(t)-\Delta J_{\rm GW}(t)|/J \ , \label{eq:deltaj}
\eeqn
where $J$ and $M$ are the ADM angular momentum and mass of the binary, respectively, 
and $M_{\rm int}(t)$ and $J_{\rm int}(t)$ are the total mass and angular momentum
of the system at time $t$ as calculated by Eqs.~\eqref{eq:Mint_sur_vol}
and \eqref{eq:Jint_sur_vol}. Note that $J_{\rm int}(0) = J$, $M_{\rm int}(0) = M$, 
and $\delta E(0)= \delta J(0) = 0$ at $t=0$. The maximum violation of energy conservation
in the simulations we present in this paper is $\delta E=2\%$ and the maximum 
angular momentum violation is $\delta J=3.4\%$.

\section{Results}
\label{sec:results}

This section presents the results from our fully relativistic
binary NS simulations with cooling. Following the merger and formation
of a HMNS remnant, which were carried out without cooling, we perform
three subsequent calculations. In one calculation cooling is never turned on
(case A). In the other two calculations cooling is
triggered at $t\approx 1600M$, at which point the HMNS remnant has relaxed to
a quasiequilibrium state -- the rest-mass density has settled and is changing on 
a secular (GW) time scale. We continue the simulations with cooling,
choosing either a short or a long cooling time scale, corresponding to case B1
and B2, respectively. Table~\ref{cases} summarizes these different
cases.

%##################################################
%          Table 2
%##################################################
\begin{table*}
\caption{Grid configurations. Here,  $N_{\rm NS}$
    denotes the number of grid points covering the smallest diameter
    of the neutron star initially (13.54km = 3.4$M$). ``Moderate'' indicates 
    the moderate resolution runs, and ``High'' the high resolution ones.}
\begin{tabular}{cccc}
  \hline
&   Grid Hierarchy (in units of $M$)$^{(a)}$
  & Max.~resolution & $N_{\rm NS}$ \\
  \hline 
Moderate  &  (181.98, 90.488, 45.244, 22.622, 15.081, 11.311, 7.541, 5.302) & $M/16.97$ & 116 \\
High &  (181.98, 90.488, 45.244, 22.622, 15.081, 11.311, 7.541, 5.043) & $M/21.22$ & 144 \\
\hline
\end{tabular}
\begin{flushleft}
$^{(a)}$ There are two sets of nested refinement boxes centered
  on each of the NSs.  This column specifies the half side length
  of the refinement boxes centered on each star. 
\end{flushleft}
\label{table:GridStructure}
\end{table*}
%##################################################

In addition, we performed all simulations 
at both moderate and high resolutions. 
The grid configurations 
are outlined in Table~\ref{table:GridStructure}. The results 
we obtained are insensitive to resolution, and for this reason
all plots that follow correspond to data from our
high-resolution runs.

\subsection{Inspiral and merger}

During inspiral and merger in case A no cooling takes place. The
evolution of the rest-mass density contours in the orbital 
plane are shown in Fig.~\ref{rho:evolution_story}. As gravitational
waves carry angular momentum away from the system, the orbits become
tighter, and the stars become strongly tidally distorted (top row, middle 
panel of Fig.~\ref{rho:evolution_story}). Shortly after the second
orbit the stars collide (top row, right 
panel of Fig.~\ref{rho:evolution_story}), marking the 
onset of the merger phase. Half an orbit later, the 
shock-heated stars become strongly sheared (bottom row, left 
panel of Fig.~\ref{rho:evolution_story}) and eventually merge and settle
in a quasiequilibrium configuration that consists of two cold cores, 
separated by hot, dense material and surrounded by a hot,
dense mantle (bottom row, middle and right panels of 
Fig.~\ref{rho:evolution_story}). The upper panel of Fig.~\ref{HMNSxz} shows 
the meridional rest-mass density contours at the same time as the
final panel of Fig.~\ref{rho:evolution_story}.
No mass outflows from the system are observed, so the HMNS has a rest mass
approximately equal to the initial rest mass of the system. This
mass now exists within an equatorial radius of about 20km and a polar
radius of about 12km. 

Fig.~\ref{K:evolution_story} shows the evolution of the orbital-plane $K = P/P_{cold }$ contours 
for case A at merger and following HMNS formation. The left panel in
Fig.~\ref{K:evolution_story} shows how the collision of the two stars begins 
to shock heat the matter. In the middle and right panels in Fig.~\ref{K:evolution_story}
the total pressure in the HMNS is clearly greater than the cold pressure
everywhere except inside the two cores. Notice the existence of a
hot area between the two cores of the remnant. In this area
$K\approx 1.5$, indicating that the thermal pressure adds a total of 50\%
additional support to the cold pressure. In the outer layers of the
remnant the thermal pressure provides up to $\sim80$\% of the total
pressure. 
The lower panel of Fig.~\ref{HMNSxz} shows the meridional 
$K$ contours. Notice that $K$ approaches $1.8$ in both the outer HMNS
layers and the hot region between the double core. These results
demonstrate that shock heating has enhanced the total pressure, which,
along with centrifugal forces, contributes to the support of the
remnant against gravitational collapse.

%%%%%%%%%%%%%%%%%%%%%%%%%%%%%%%%%%%%%%%%%%%%%%%%%%%%%%%%%%%%%%%%%%%%%%%%%%%%%%%%%%%%%%%%%%%%%%%%%%
%                                           						Figure 5
%%%%%%%%%%%%%%%%%%%%%%%%%%%%%%%%%%%%%%%%%%%%%%%%%%%%%%%%%%%%%%%%%%%%%%%%%%%%%%%%%%%%%%%%%%%%%%%%%%
\begin{figure}%[h]
\centering
\subfigure{\includegraphics[width=0.445\textwidth]{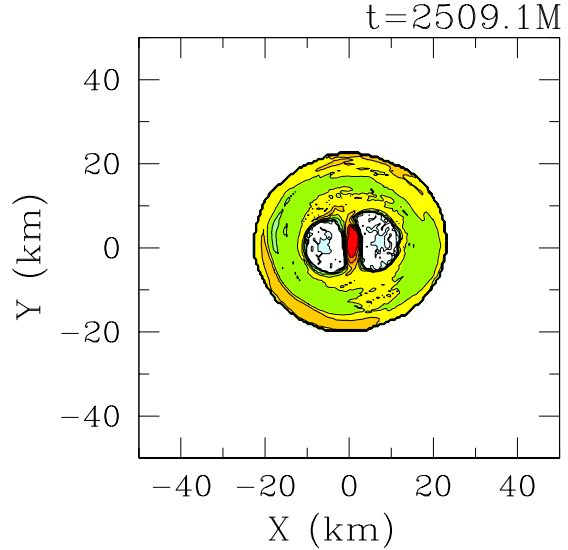}}
\caption{Case A orbital-plane temperature contours. Contours are plotted according to ${\rm T} = {\rm T}_{\rm max} 10^{-0.136j}$,
  ($j$=0, 1, ... 8), where ${\rm T}_{\rm max} = 2.57 \times 10^{11} {\rm K} = 22.18$MeV. The color coding here is the same as
 in Fig.~\ref{rho:evolution_story}. A density cutoff of $10^{-2}\rho_{0,max}$ has been imposed, where
 $\rho_{0,{\rm max}}$ is the maximum density on the grid.
The maximum temperature is at the center of the HMNS remnant. The rms temperature in the remnant is 
$\bar {\rm T} = 6.35\times 10^{10} {\rm K} \approx 5.5\rm$MeV. 
Here $M=1.32\times 10^{-5}$s$=3.98$km is the ADM mass.}
\label{HMNStemp}
\centering
\end{figure}
%%%%%%%%%%%%%%%%%%%%%%%%%%%%%%%%%%%%%%%%%%%%%%%%%%%%%%%%%%%%%%%%%%%%%%%%%%%%%%%%%%%%%%%%%%%%%%%%%%
%%%%%%%%%%%%%%%%%%%%%%%%%%%%%%%%%%%%%%%%%%%%%%%%%%%%%%%%%%%%%%%%%%%%%%%%%%%%%%%%%%%%%%%%%%%%%%%%%%

Following HMNS formation at about $t\approx 500M = 6.6(M/2.69M_\odot)
\rm ms$, the remnant survives for a long quasistationary epoch, during
which the maximum density increases almost linearly with time (see
left panel of Fig.\ref{fig:ang_vs_rhoc}). Similar behavior is reported
in \cite{2008PhRvD..78h4033B} when using the same $\Gamma$-law EOS
adopted here and no cooling.% [$P=(\Gamma - 1) \rho_0\epsilon$]. 

In addition, \cite{2008PhRvD..78h4033B} performed a simulation of the same
system, but with a strict {\it polytropic} EOS ($P=\kappa \rho_0^{\Gamma}$),
in which shocks are artificially suppressed. In this case, it is found that
the resulting HMNS collapsed when $\rho_{0,\rm max} \approx 2
\rho_{0,\rm max,initial}$ at which point $t\approx 21ms$. Applying this same density criterion to the
$\Gamma$-law EOS, they extrapolate that the HMNS would collapse at
$t\approx 110(M/2.69M_\odot) \rm ms$. Our simulations show that
$\rho_{0,\rm max} \approx 2 \rho_{0,\rm max,initial}$ at $t\approx 105(M/2.69M_\odot) \rm
ms$, in good agreement with \cite{2008PhRvD..78h4033B}'s result.
In a follow-up calculation \cite{Rezzolla:2010fd}, the same authors
demonstrate that the HMNS remnant collapses to a BH at
$t\approx 130(M/2.69M_\odot)$ms. 

In the polytropic EOS simulation (shocks disallowed), centrifugal forces provide the only source of 
support against collapse and $t_{\rm coll} \sim t_{\rm GW}$. However, 
in the $\Gamma$-law EOS simulation (shocks allowed), 
additional pressure from thermal support is also present, so
$t_{\rm coll}$ can be larger. In fact, in the absence of cooling, a sufficiently hot remnant may {\it never}
collapse. This is what \cite{2011PhRvL.107e1102S} conclude from their
case M simulation, as they demonstrated that a hot TOV star 
could support in equilibrium a mass greater than the total mass 
of their merged remnant. 

%%%%%%%%%%%%%%%%%%%%%%%%%%%%%%%%%%%%%%%%%%%%%%%%%%%%%%%%%%%%%%%%%%%%%%%%%%%%%%%%%%%%%%%%%%%%%%%%%%
%                                           						Figure 6
%%%%%%%%%%%%%%%%%%%%%%%%%%%%%%%%%%%%%%%%%%%%%%%%%%%%%%%%%%%%%%%%%%%%%%%%%%%%%%%%%%%%%%%%%%%%%%%%%%
\begin{figure*}[th]
\centering
\subfigure{\includegraphics[width=0.325\textwidth]{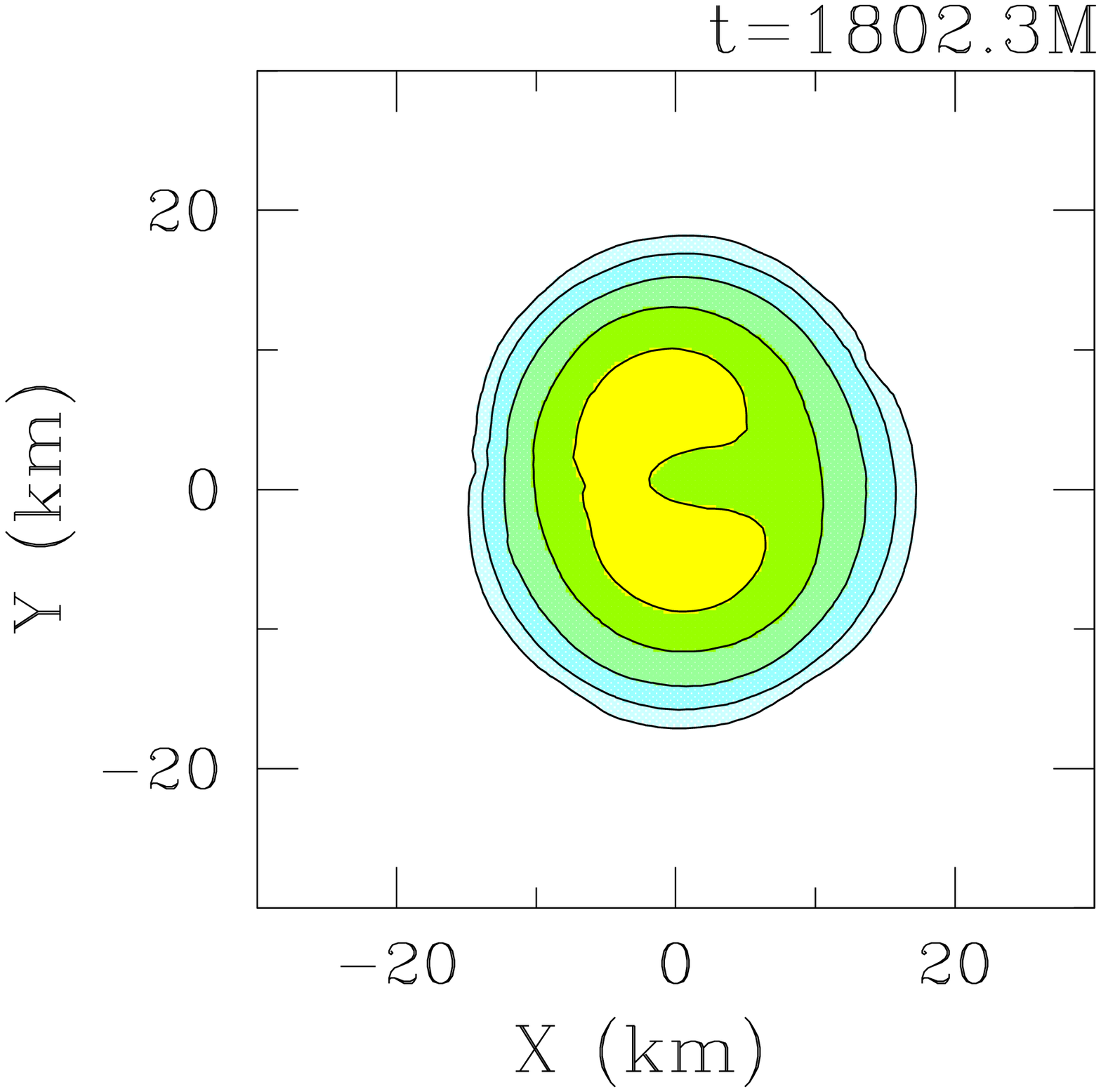}}
\subfigure{\includegraphics[width=0.325\textwidth]{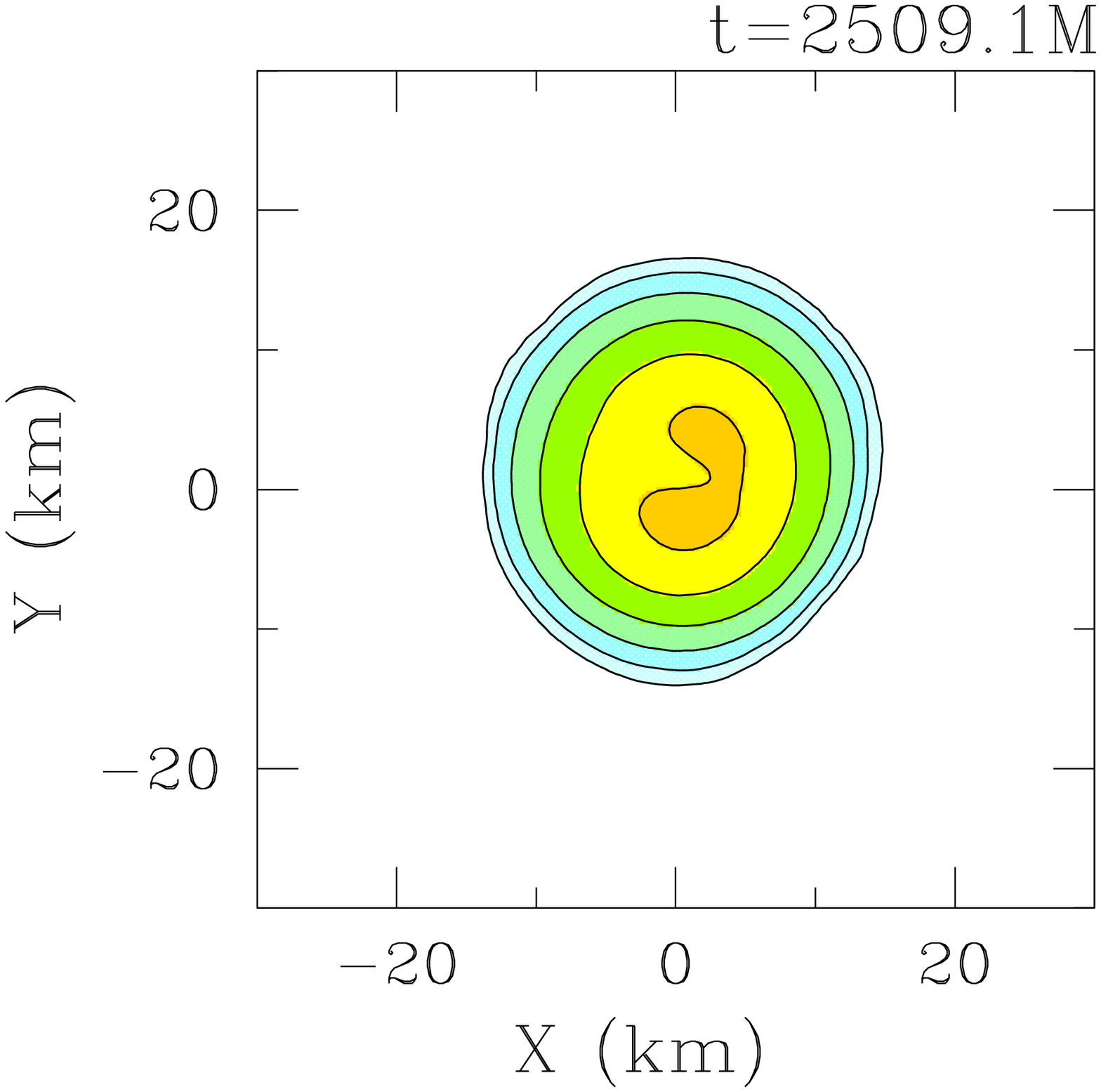}}
\subfigure{\includegraphics[width=0.325\textwidth]{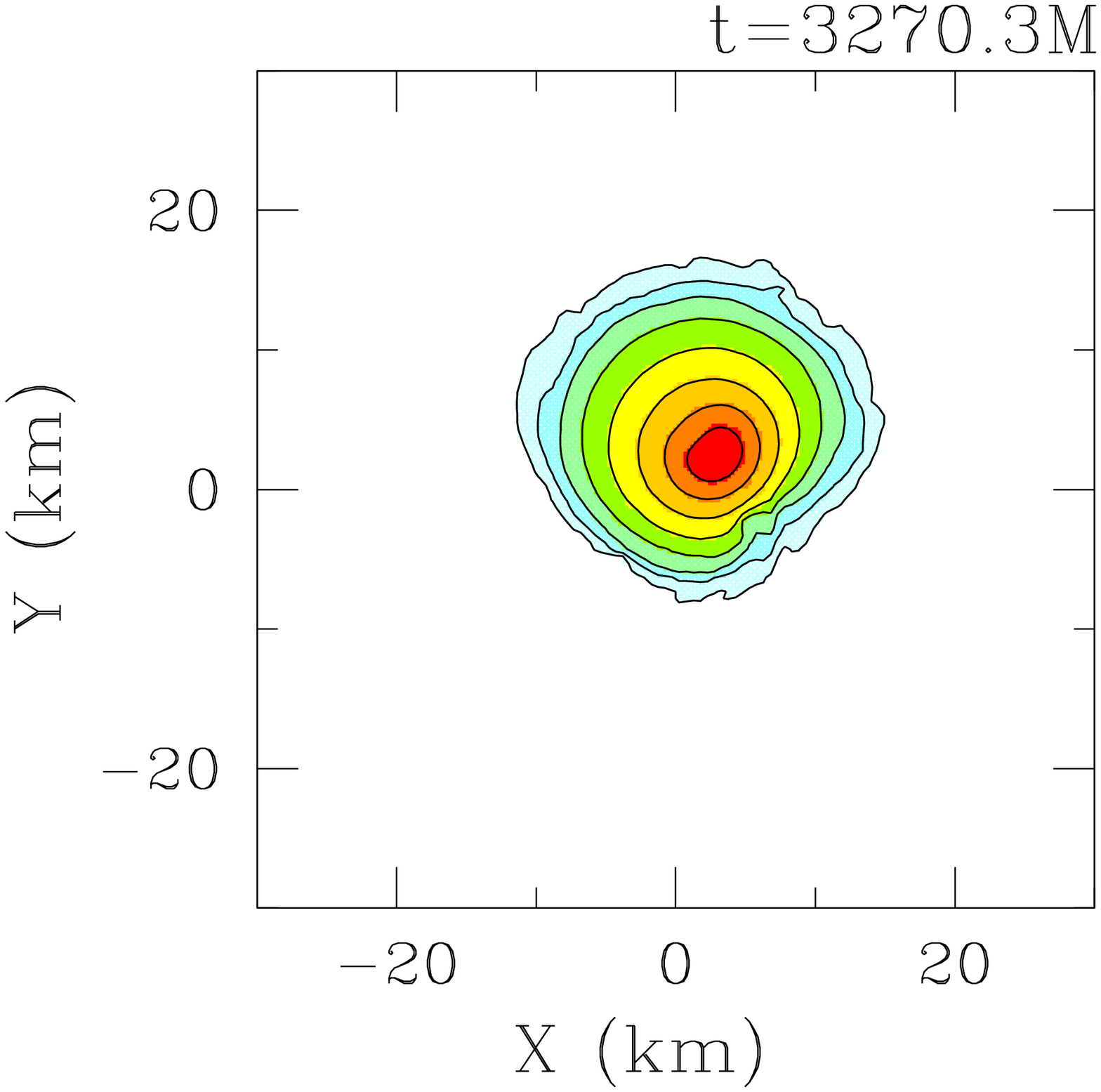}}
\subfigure{\includegraphics[width=0.325\textwidth]{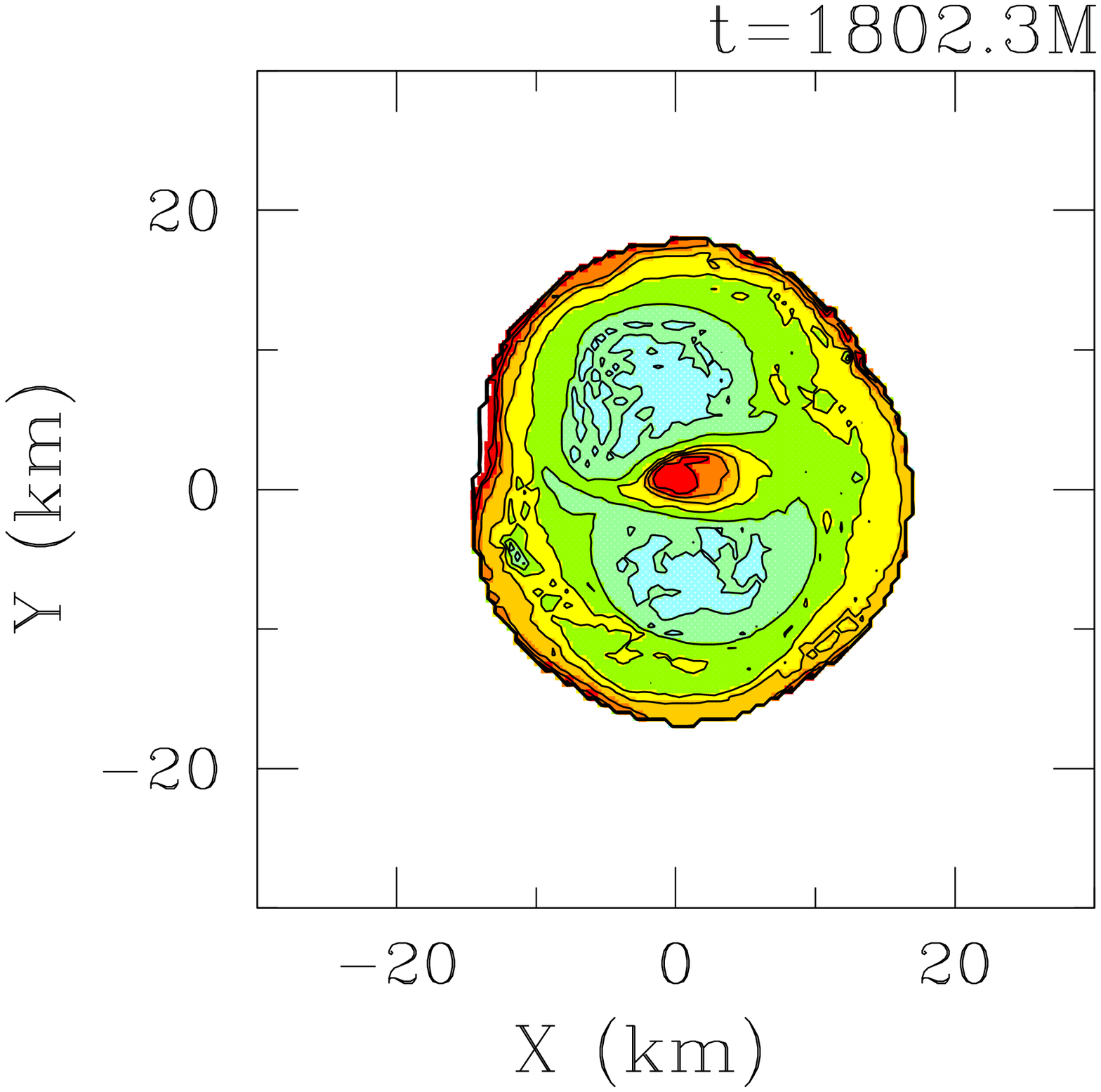}}
\subfigure{\includegraphics[width=0.325\textwidth]{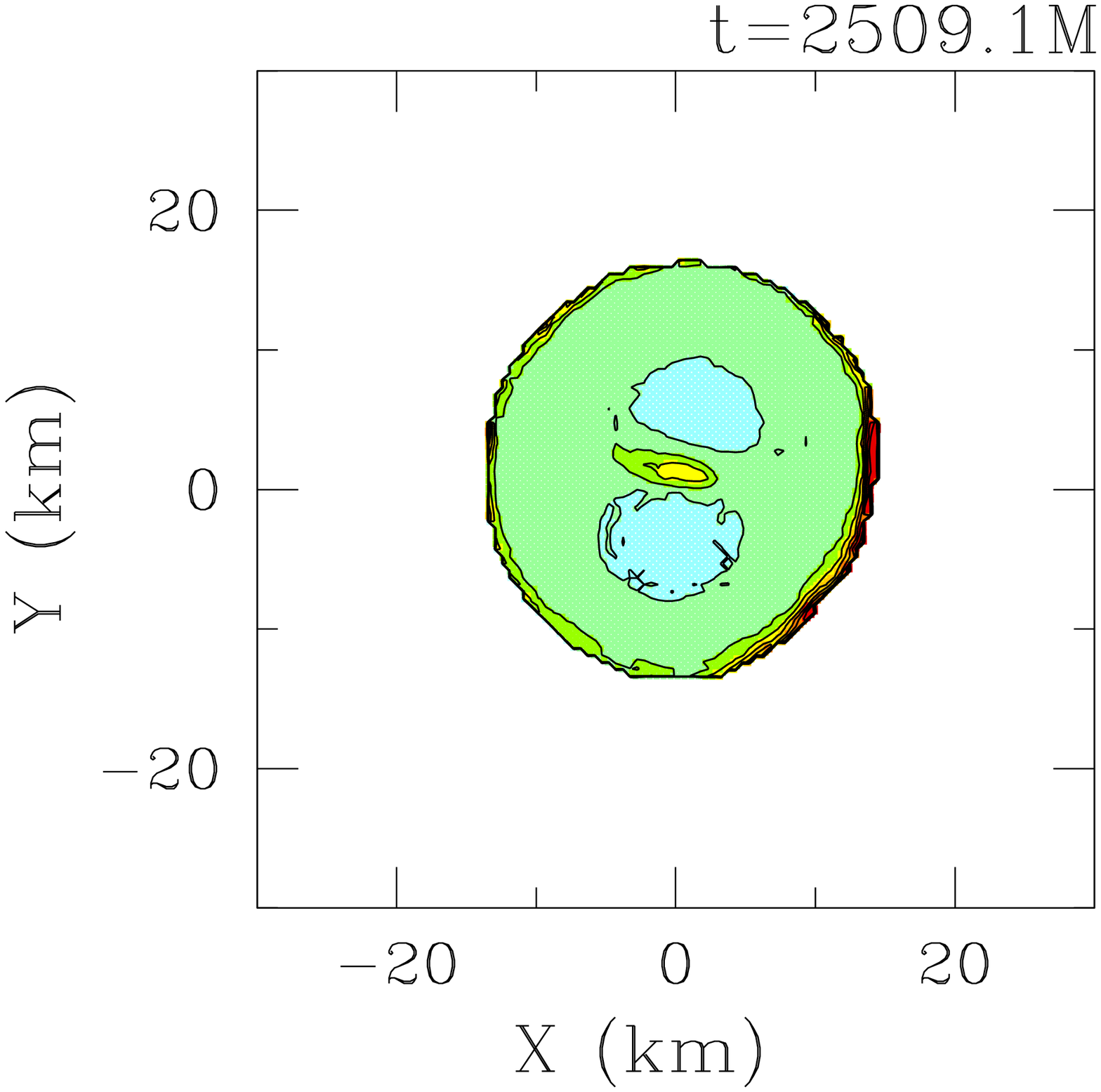}}
\subfigure{\includegraphics[width=0.325\textwidth]{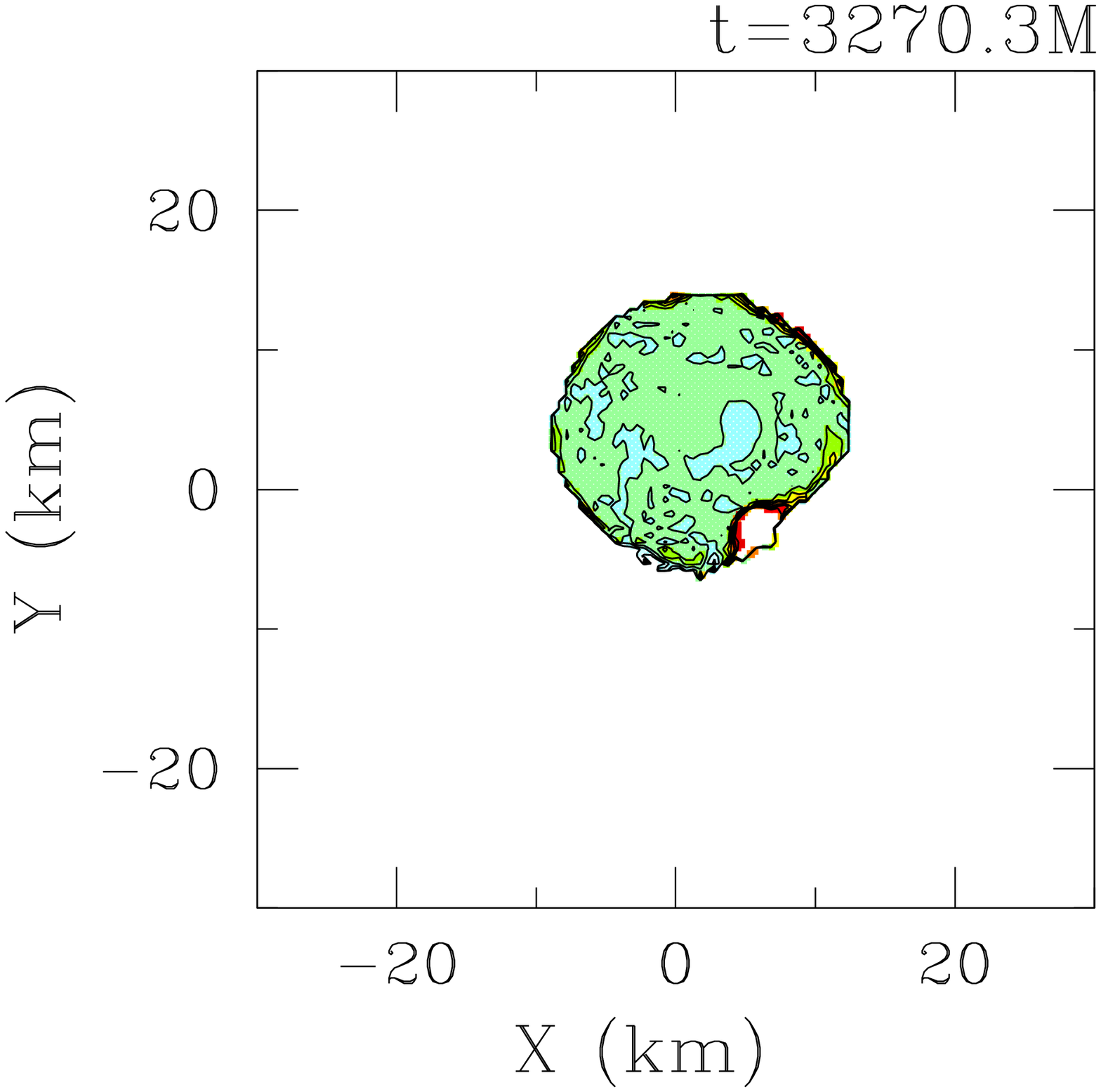}}
\caption{Upper row: Case B2 orbital-plane rest-mass density contours at selected times. Contours are plotted according to
 $\rho_0 =  \rho_{0,{\rm max}} (10^{-0.2j+0.568})$, ($j$=0, 1, ... 8), where $\kappa \rho_{0,{\rm max}} = 0.0917$, 
or $\rho_{0,{\rm max}}=4.58\times 10^{14}\mbox{g cm}^{-3}(1.45M_\odot/M_0)^2$. 
Lower row: Case B2 orbital-plane $K$ contours at selected times. Contours are plotted according to $K = K_{\rm max}10^{-0.025j}$,
($j$=0, 1, ... 8). Here $K_{\rm max} = 1.6$. The color coding is the same as used in Fig.~\ref{rho:evolution_story}, 
with light blue indicating to $K\approx 1$, yellow $K\approx 1.2$ and dark red $K\approx1.4$. 
Here $M=1.32\times 10^{-5} (M_0/1.45M_\odot)$s$=3.98(M_0/1.45M_\odot)$km is the ADM mass, and $M_0$ denotes
the rest mass of each star.
\label{fig:K_rho_evolution_with_cooling}
}
\centering
\end{figure*}
%%%%%%%%%%%%%%%%%%%%%%%%%%%%%%%%%%%%%%%%%%%%%%%%%%%%%%%%%%%%%%%%%%%%%%%%%%%%%%%%%%%%%%%%%%%%%%%%%%
%%%%%%%%%%%%%%%%%%%%%%%%%%%%%%%%%%%%%%%%%%%%%%%%%%%%%%%%%%%%%%%%%%%%%%%%%%%%%%%%%%%%%%%%%%%%%%%%%%
  
Given that the $\Gamma$-law EOS remnant collapses about 
100 ms later than the polytropic one, it is clear that shocks
play a key dynamical role. However, shocks, which occur on a hydrodynamical time scale, 
give rise to at least three effects:
\begin{enumerate}
\item  they heat the gas, increasing the total pressure support;

\item  they affect the matter profile;

\item  they redistribute the angular momentum.

\end{enumerate}
{\it A priori} it is not clear which of these effects is responsible 
for prolonging the HMNS lifetime, and the answer cannot be determined 
by comparing simulations that do not allow for shocks to those that do. 
%For example, one cannot claim based on such a comparison that the HMNS lifetime is necessarily 
%prolonged due to shock-induced thermal support alone. After all, the shock-influenced
%angular momentum and matter profiles can change both the GW time scale and the amount 
%of centrifugal support, which may prolong the HMNS lifetime.

To investigate the importance of the shock-induced thermal pressure support against collapse, cooling 
the hot HMNS remnant is crucial. But what is the neutrino cooling time scale?

%%%%%%%%%%%%%%%%%%%%%%%%%%%%%%%%%%%%%%%%%%%%%%%%%%%%%%%%%%%%%%%%%%%%%%%%%%%%%%%%%%%%%%%%%%%%%%%%%%
%                                           						Figure 7
%%%%%%%%%%%%%%%%%%%%%%%%%%%%%%%%%%%%%%%%%%%%%%%%%%%%%%%%%%%%%%%%%%%%%%%%%%%%%%%%%%%%%%%%%%%%%%%%%%
\begin{figure}[t]
\centering
\includegraphics[width=0.445\textwidth,angle=0]{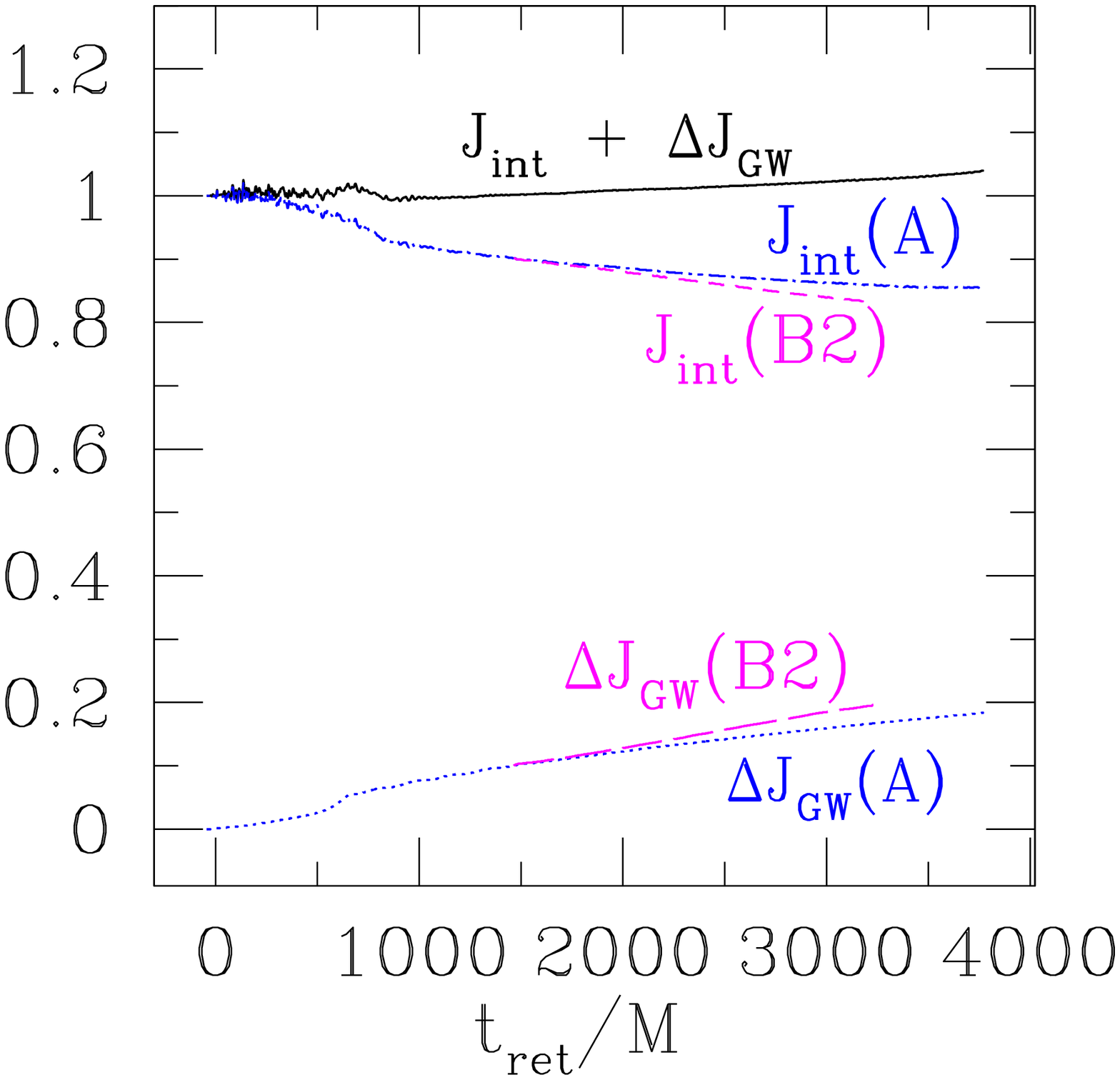}
\caption{Angular momentum vs time. Here $J_{\rm int}(A)$ and $\Delta J_{\rm GW}(A)$ are the total
 angular momentum and the angular momentum carried away by GWs, respectively,
for case A. $J_{\rm int}(B2)$ and $\Delta J_{\rm GW}(B2)$ correspond to case B2. 
The sum in both cases is the same and given by $J_{\rm int}+\Delta J_{\rm GW}$.
All quantities are normalized by the ADM angular momentum of the binary $J$.
\label{fig:total_ang}
}
\centering
\end{figure}
%%%%%%%%%%%%%%%%%%%%%%%%%%%%%%%%%%%%%%%%%%%%%%%%%%%%%%%%%%%%%%%%%%%%%%%%%%%%%%%%%%%%%%%%%%%%%%%%%%
%%%%%%%%%%%%%%%%%%%%%%%%%%%%%%%%%%%%%%%%%%%%%%%%%%%%%%%%%%%%%%%%%%%%%%%%%%%%%%%%%%%%%%%%%%%%%%%%%%

Analyzing the nascent HMNS in case A, we can now estimate a realistic
cooling time scale directly from our simulation data. 
In Fig.~\ref{HMNStemp}
we show the case A orbital-plane temperature contours of the HMNS remnant. 
The maximum and rms temperatures are  $\sim 22$MeV and
$\sim 5.5$MeV, respectively. Using these values for the neutrino
energy, and setting $M=2.7M_\odot$ and $R=16 \rm km$, Eq.\eqref{tCool}
yields a neutrino diffusion time scale of $\sim 165{\rm ms}-2.64\rm
s$. Note that this range is consistent with our discussion in
Sec.~\ref{sec:timescales}, where the neutrino diffusion time scale
was estimated to be comparable to the gravitational wave time scale of $\sim 120\rm ms$.

Therefore, as the additional thermal pressure is a dynamically important source of support, 
then cooling must be incorporated in numerical simulations
to accurately determine the time interval between merger 
and delayed collapse. Low energy neutrinos ($E_\nu \lesssim 10$MeV), that 
escape the HMNS on a time scale $\lesssim t_{\rm GW}$, can remove a sufficiently large amount of 
thermal energy to accelerate the collapse. This is the main point we want to emphasize in this paper.
Our conclusion may even be more important when a stiff EOS is employed, as the HMNS 
will be less compact resulting in less effective GW emission. 

\subsection{Cooling study}

To assess the impact of the shock-induced thermal pressure in the
HMNS remnant, we slowly remove the thermal pressure using our
covariant cooling technique.
We chose two cooling time scales for our study. The short one is 
$\tau_{c,1} = 6.5 t_{\rm d}$ for case B1 and the long one is $\tau_{c,2} = 2\tau_{c,1}=13 t_{\rm d}$
for case B2, where 
\labeq{}{
t_{\rm d} \equiv \frac{1}{\sqrt{\bar \rho}} \approx 23 M. 
}
is the dynamical time scale of the HMNS, where $\bar \rho$ is the mean HMNS density. 
This choice is arbitrary, but it is set 
so that the cooling time scale is significantly longer than the dynamical time scale of the remnant, as in 
physically realistic stars, but short enough for our simulations to be completed within a reasonable
time.

Our runs with cooling lead to BH formation within a few cooling time scales. 
This can be seen in the left two panels of Fig.~\ref{fig:ang_vs_rhoc}
where we plot the maximum rest-mass density $\rho_{0,\rm max}$ and the
minimum of the lapse function $\alpha_{\rm min}$ versus time,
respectively. From these plots it becomes clear that case A does not
collapse within the integration time, while both cases B1 and B2 form a BH. 
Notice that in
cases B1 and B2, when $\rho_{0,\rm max}$ roughly equals two
times its initial value, the lapse function collapses and a BH
forms. Note that this is consistent with the polytropic runs of
\cite{2008PhRvD..78h4033B}, in which shocks were suppressed. 
Therefore, in all cases with the adopted EOS,
collapse takes place when $\rho_{0,\rm max} \approx 2 \rho_{0,\rm max,initial}$, 
when thermal energy is drained from the system.

If the collapse is driven by cooling, then we naturally 
expect that a longer cooling time scale 
will increase the lifetime of the HMNS. This is precisely 
what we find: the collapse in case
B2 occurs later than in case B1. Further evidence of cooling-induced
collapse is shown in the right panel of
Fig.~\ref{fig:ang_vs_rhoc}. There the tracks of $\rho_{0,\rm max}$
against the total angular momentum of the remnant $J_{\rm int}$ are plotted using Eq.~\eqref{eq:Jint_sur_vol}. 
This plot demonstrates that during the post-merger evolution for the same $J_{\rm int}$, $\rho_{0,\rm max}$ 
is always larger with cooling present (e.g. $\rho_{0,\rm max}$ in cases B1, B2 is
$\sim 30\%$ larger than in case A for the smallest $J_{\rm int}$ reached in case A). 
Therefore, it is the reduction of thermal pressure that leads 
to a more compact remnant and not angular momentum loss driven by GWs.

In Fig.~\ref{fig:K_rho_evolution_with_cooling} we show the evolution of the 
rest-mass density (upper row) and $K$ contours (lower row) for case B2. The 
selected times correspond to $1$, $8$, and $11$ cooling time scales after cooling was turned on \footnote{The
corresponding plots in case B1 are similar and therefore not shown.}. The right
panel corresponds to a time shortly before an apparent horizon forms. 
These plots indicate that as thermal pressure is removed, the double
cores approach one another and the HMNS becomes more compact
(notice the density increase and the shrinking size of the remnant with increasing time). 
We find that when the {\it hot} area between the two cores is cooled to 
$K\approx 1.05$ the two cores merge and form a single-core HMNS. Shortly after 
this occurs (at $t\approx 3000M$) the remnant undergoes catastrophic collapse.

Based on these results we conclude that thermal pressure contributes significantly to 
support against collapse.

\subsection{Angular momentum conservation}

Figure~\ref{fig:total_ang} plots the 
evolution of total angular momentum ($J_{\rm int}$) and angular
momentum carried off by GWs ($\Delta J_{\rm GW}$), normalized by the ADM
angular momentum of the binary ($J$) for cases A and B2. 
We find that $(J_{\rm int}+\Delta J_{\rm GW})/J$, which should be equal to unity at all times,
is the same and close to unity for all three cases A, B1 and B2. 
This implies that cooling carries off negligible amounts of angular momentum
which is consistent with earlier estimates \cite{Baumgarte:1998sn}.
The maximum violation of angular momentum conservation is 
$\delta J \approx 3.4\%$. Notice that $J_{\rm int}$ is smaller in case B2 than in case A, 
while $\Delta J_{\rm GW}$ is larger in case B2 than in case A.
The distinction is due to the fact that as thermal energy is radiated away the remnant 
becomes more compact, enabling GWs to remove angular momentum 
faster. We conclude that cooling accelerates the collapse of the HMNS 
by the combined action of two effects: 
\begin{enumerate}
\item Cooling removes thermal pressure support, yielding a more
  compact remnant.
\item As the remnant becomes more compact, GWs are able to carry away
  angular momentum more efficiently.
\end{enumerate}

\section{Summary and Future work}
\label{sec:summaryandfuturework}

A differentially rotating, quasiequilibrium HMNS is a transient configuration that 
can arise following the merger of NSNS binaries. The mass of a HMNS is larger than 
the maximum mass that can be supported by a cold EOS, even with
maximal uniform rotation. A HMNS will eventually 
undergo ``delayed collapse'' on a secular (dissipative) time scale and may 
power a sGRB. 

When HMNSs are born in NSNS mergers, they are rapidly differentially rotating 
and {\it hot} due to shock heating. Therefore, HMNSs will collapse 
to a BH either on an angular momentum loss/magnetic braking time scale
or on a cooling time scale. A priori it is not clear which of the two above mechanisms is 
most important for holding up an HMNS against collapse: 
{\it centrifugal forces or thermal pressure}. The answer to this question is
still open and may depend on the stellar model, EOS, and initial
magnetic fields.

Determining which mechanism drives a  HMNS to collapse
has observational consequences;
the time scale of collapse will set the interval between the NSNS
merger chirp signal and the delayed collapse burst signal,
which may be measured by LIGO/VIRGO. Careful modeling of HMNS physics
will thus place constraints on magnetic field magnitudes, the
existence of bar modes, and/or the relevant cooling mechanisms. In
addition, such observations could place constraints on the temperature
of matter as well as the nuclear EOS. 

To disentangle the effects of thermal support from those of rotational support,
previous studies compared results from NSNS 
simulations that suppress shocks to those that allow shocks.
If the HMNS remnant lives longer in the case with
shocks than without, then it is tempting to infer
that thermal pressure due to shock heating is solely responsible.
However it is not possible to draw such a firm conclusion because shocks,
which act on a hydrodynamical time scale, not only heat the gas,
thereby increasing the total pressure support, but also affect the
matter profile and redistribute angular momentum. Different matter and angular
momentum profiles alone can increase the lifetime of a HMNS via an increase of both the
GW time scale and the amount of differential rotational support.

To address this issue, we first performed long-term, high-resolution GRHD 
NSNS simulations through inspiral, merger, and
HMNS formation, allowing for shocks. 
Following HMNS formation, we continue the evolution
both with and without cooling. When cooling is
turned off, the remnant collapses on the GW time scale.
However, when cooling is turned on we find that
the HMNS collapses and forms a BH within a few cooling time scales.

Our simulations demonstrate that shock-induced thermal pressure 
is a {\it significant} source of support against
gravitational collapse in the case of a stiff $\Gamma$-law EOS -- a result
consistent with simulations that employ a more realistic
EOS~\cite{2011PhRvL.107e1102S} -- and show explicitly
that cooling can induce the catastrophic collapse of a HMNS. 
Estimating the temperature of the HMNS remnant,
we find that a realistic neutrino cooling time scale is of order a few
$100$ms. Given that the estimated cooling and angular momentum
loss/magnetic braking time scales can be comparable, 
cooling should be accounted for to accurately determine the
lifetime of a HMNS. Therefore simulations that implement cooling will
lead to earlier collapse than simulations that ignore it, otherwise
the predicted GW and EM signatures from these delayed collapse events may be incorrect.

Therefore, to accurately determine the
lifetime of HMNS remnants, neutrino cooling physics should be
incorporated in NSNS simulations.
In the future we plan to revisit the subject using a more realistic
neutrino leakage scheme, such as that used in
\cite{Sekiguchi:2010fh,2011PhRvL.107e1102S}, in conjunction with more
realistic treatment of the microphysics involved.

\acknowledgments

The authors would like to thank B. Giacomazzo, Y. T. Liu, and Y. Sekiguchi
for helpful discussions. This paper was supported in part by NSF Grants AST-1002667, and
PHY-0963136 as well as NASA Grant NNX11AE11G at the University of Illinois at Urbana-Champaign.

%%%%%%%%%%%%%%
% APPENDICES %
%%%%%%%%%%%%%%

\bibliography{paper}

%Merlin.mbs v4.21 2009-07-09.
\begin{thebibliography}{10}%
\makeatletter
\providecommand \@ifxundefined [1]{%
 \ifx #1\undefined \expandafter \@firstoftwo
 \else \expandafter \@secondoftwo
\fi
}%
\providecommand \@ifnum [1]{%
 \ifnum #1\expandafter \@firstoftwo
 \else \expandafter \@secondoftwo
\fi
}%
\providecommand \enquote [1]{``#1''}%
\providecommand \bibnamefont  [1]{#1}%
\providecommand \bibfnamefont [1]{#1}%
\providecommand \citenamefont [1]{#1}%
\providecommand\href[0]{\@sanitize\@href}%
\providecommand\@href[1]{\endgroup\@@startlink{#1}\endgroup\@@href}%
\providecommand\@@href[1]{#1\@@endlink}%
\providecommand \@sanitize [0]{\begingroup\catcode`\&12\catcode`\#12\relax}%
\@ifxundefined \pdfoutput {\@firstoftwo}{%
 \@ifnum{\z@=\pdfoutput}{\@firstoftwo}{\@secondoftwo}%
}{%
 \providecommand\@@startlink[1]{\leavevmode\special{html:<a href="#1">}}%
 \providecommand\@@endlink[0]{\special{html:</a>}}%
}{%
 \providecommand\@@startlink[1]{%
  \leavevmode
  \pdfstartlink
   attr{/Border[0 0 1 ]/H/I/C[0 1 1]}%
   user{/Subtype/Link/A<</Type/Action/S/URI/URI(#1)>>}%
  \relax
 }%
 \providecommand\@@endlink[0]{\pdfendlink}%
}%
\providecommand \url  [0]{\begingroup\@sanitize \@url }%
\providecommand \@url [1]{\endgroup\@href {#1}{\urlprefix}}%
\providecommand \urlprefix [0]{URL }%
\providecommand \Eprint[0]{\href }%
\@ifxundefined \urlstyle {%
  \providecommand \doi [1]{doi:\discretionary{}{}{}#1}%
}{%
  \providecommand \doi [0]{doi:\discretionary{}{}{}\begingroup
  \urlstyle{rm}\Url }%
}%
\providecommand \doibase [0]{http://dx.doi.org/}%
\providecommand \Doi[1]{\href{\doibase#1}}%
\providecommand \bibAnnote [3]{%
  \BibitemShut{#1}%
  \begin{quotation}\noindent
    \textsc{Key:}\ #2\\\textsc{Annotation:}\ #3%
  \end{quotation}%
}%
\providecommand \bibAnnoteFile [2]{%
  \IfFileExists{#2}{\bibAnnote {#1} {#2} {\input{#2}}}{}%
}%
\providecommand \typeout [0]{\immediate \write \m@ne }%
\providecommand \selectlanguage [0]{\@gobble}%
\providecommand \bibinfo [0]{\@secondoftwo}%
\providecommand \bibfield [0]{\@secondoftwo}%
\providecommand \translation [1]{[#1]}%
\providecommand \BibitemOpen[0]{}%
\providecommand \bibitemStop [0]{}%
\providecommand \bibitemNoStop [0]{.\EOS\space}%
\providecommand \EOS [0]{\spacefactor3000\relax}%
\providecommand \BibitemShut [1]{\csname bibitem#1\endcsname}%
%</preamble>
\bibitem{LIGO1}%
  \BibitemOpen
  \bibfield{author}{%
  \bibinfo {author} {\bibfnamefont{B.}~\bibnamefont{{Abbott}}}\ and\ \bibinfo
  {author} {\bibnamefont{{the LIGO Scientific Collaboration}}},\ }%
  \bibfield{journal}{%
  \Doi{10.1103/PhysRevD.77.062002}{\bibinfo {journal} {\prd}}\ }%
  \textbf{\bibinfo {volume} {77}},\ \bibinfo {pages} {062002} (\bibinfo {month}
  {Mar.}\ \bibinfo {year} {2008})%
  \bibAnnoteFile{NoStop}{LIGO1}%
\bibitem{LIGO2}%
  \BibitemOpen
  \bibfield{author}{%
  \bibinfo {author} {\bibfnamefont{D.~A.}\ \bibnamefont{{Brown}}}, \bibinfo
  {author} {\bibfnamefont{S.}~\bibnamefont{{Babak}}}, \bibinfo {author}
  {\bibfnamefont{P.~R.}\ \bibnamefont{{Brady}}}, \bibinfo {author}
  {\bibfnamefont{N.}~\bibnamefont{{Christensen}}}, \bibinfo {author}
  {\bibfnamefont{T.}~\bibnamefont{{Cokelaer}}}, \bibinfo {author}
  {\bibfnamefont{J.~D.~E.}\ \bibnamefont{{Creighton}}}, \bibinfo {author}
  {\bibfnamefont{S.}~\bibnamefont{{Fairhurst}}}, \bibinfo {author}
  {\bibfnamefont{G.}~\bibnamefont{{Gonzalez}}}, \bibinfo {author}
  {\bibfnamefont{E.}~\bibnamefont{{Messaritaki}}}, \bibinfo {author}
  {\bibfnamefont{B.~S.}\ \bibnamefont{{Sathyaprakash}}}, \bibinfo {author}
  {\bibfnamefont{P.}~\bibnamefont{{Shawhan}}},\ and\ \bibinfo {author}
  {\bibfnamefont{N.}~\bibnamefont{{Zotov}}},\ }%
  \bibfield{journal}{%
  \bibinfo {journal} {Class.~Quant.~Grav.}\ }%
  \textbf{\bibinfo {volume} {21}},\ \bibinfo {pages} {S1625} (\bibinfo {month}
  {Oct.}\ \bibinfo {year} {2004})%
  \bibAnnoteFile{NoStop}{LIGO2}%
\bibitem{VIRGO1}%
  \BibitemOpen
  \bibfield{author}{%
  \bibinfo {author} {\bibfnamefont{F.}~\bibnamefont{{Acernese}}}\ and\ \bibinfo
  {author} {\bibnamefont{{the VIRGO Collaboration}}},\ }%
  \bibfield{journal}{%
  \Doi{10.1088/0264-9381/23/19/S01}{\bibinfo {journal} {Class.~Quant.~Grav.}}\
  }%
  \textbf{\bibinfo {volume} {23}},\ \bibinfo {pages} {S635} (\bibinfo {month}
  {Oct.}\ \bibinfo {year} {2006})%
  \bibAnnoteFile{NoStop}{VIRGO1}%
\bibitem{VIRGO2}%
  \BibitemOpen
  \bibfield{author}{%
  \bibinfo {author} {\bibfnamefont{F.}~\bibnamefont{{Beauville}}}\ and\
  \bibinfo {author} {\bibnamefont{{the LIGO-VIRGO Working Group}}},\ }%
  \bibfield{journal}{%
  \Doi{10.1088/0264-9381/25/4/045001}{\bibinfo {journal} {Classical and Quantum
  Gravity}}\ }%
  \textbf{\bibinfo {volume} {25}},\ \bibinfo {pages} {045001} (\bibinfo {month}
  {Feb.}\ \bibinfo {year} {2008})%
  \bibAnnoteFile{NoStop}{VIRGO2}%
\bibitem{GEO}%
  \BibitemOpen
  \bibfield{author}{%
  \bibinfo {author} {\bibfnamefont{H.}~\bibnamefont{{L{\"u}ck}}}\ and\ \bibinfo
  {author} {\bibnamefont{{the GEO600 collaboration}}},\ }%
  \bibfield{journal}{%
  \Doi{10.1088/0264-9381/23/8/S10}{\bibinfo {journal} {Class.~Quant.~Grav.}}\
  }%
  \textbf{\bibinfo {volume} {23}},\ \bibinfo {pages} {S71} (\bibinfo {month}
  {Apr.}\ \bibinfo {year} {2006})%
  \bibAnnoteFile{NoStop}{GEO}%
\bibitem{KAGRA}%
  \BibitemOpen
  \bibfield{author}{%
  \bibinfo {author} {\bibfnamefont{K.}~\bibnamefont{Somiya}} (\bibinfo
  {collaboration} {for the LCGT Collaboration})}%
   (\bibinfo {year} {2011}),\
  \Eprint{http://arxiv.org/abs/1111.7185}{arXiv:1111.7185 [gr-qc]}%
  \bibAnnoteFile{NoStop}{KAGRA}%
%%CITATION = ARXIV:1111.7185;%%
\bibitem{NGO}%
  \BibitemOpen
  \bibfield{author}{%
  \bibinfo {author} {\bibfnamefont{P.}~\bibnamefont{Amaro-Seoane}}, \bibinfo
  {author} {\bibfnamefont{S.}~\bibnamefont{Aoudia}}, \bibinfo {author}
  {\bibfnamefont{S.}~\bibnamefont{Babak}}, \bibinfo {author}
  {\bibfnamefont{P.}~\bibnamefont{Binetruy}}, \bibinfo {author}
  {\bibfnamefont{E.}~\bibnamefont{Berti}}, \emph{et~al.}}%
   (\bibinfo {year} {2012}),\
  \Eprint{http://arxiv.org/abs/1202.0839}{arXiv:1202.0839 [gr-qc]}%
  \bibAnnoteFile{NoStop}{NGO}%
%%CITATION = ARXIV:1202.0839;%%
\bibitem{DECIGO}%
  \BibitemOpen
  \bibfield{author}{%
  \bibinfo {author} {\bibfnamefont{S.}~\bibnamefont{{Kawamura}}}\ and\ \bibinfo
  {author} {\bibnamefont{{the DECIGO collaboration}}},\ }%
  \bibfield{journal}{%
  \Doi{10.1088/0264-9381/23/8/S17}{\bibinfo {journal} {Class.~Quant.~Grav.}}\
  }%
  \textbf{\bibinfo {volume} {23}},\ \bibinfo {pages} {S125} (\bibinfo {month}
  {Apr.}\ \bibinfo {year} {2006})%
  \bibAnnoteFile{NoStop}{DECIGO}%
\bibitem{BSBook}%
  \BibitemOpen
  \bibfield{author}{%
  \bibinfo {author} {\bibfnamefont{T.~W.}\ \bibnamefont{{Baumgarte}}}\ and\
  \bibinfo {author} {\bibfnamefont{S.~L.}\ \bibnamefont{{Shapiro}}},\ }%
  \emph{\bibinfo {title} {{Numerical Relativity: Solving Einstein's Equations
  on the Computer}}}\ (\bibinfo {publisher} {Cambridge University Press},\
  \bibinfo {year} {2010})%
  \bibAnnoteFile{NoStop}{BSBook}%
\bibitem{Hinder:2010vn}%
  \BibitemOpen
  \bibfield{author}{%
  \bibinfo {author} {\bibfnamefont{I.}~\bibnamefont{Hinder}},\ }%
  \bibfield{journal}{%
  \Doi{10.1088/0264-9381/27/11/114004}{\bibinfo {journal} {Class.Quant.Grav.}}\
  }%
  \textbf{\bibinfo {volume} {27}},\ \bibinfo {pages} {114004} (\bibinfo {year}
  {2010}),\ \Eprint{http://arxiv.org/abs/1001.5161}{arXiv:1001.5161 [gr-qc]}%
  \bibAnnoteFile{NoStop}{Hinder:2010vn}%
%%CITATION = ARXIV:1001.5161;%%
\bibitem{BNSlr}%
  \BibitemOpen
  \bibfield{author}{%
  \bibinfo {author} {\bibfnamefont{J.~A.}\ \bibnamefont{Faber}}\ and\ \bibinfo
  {author} {\bibfnamefont{F.~A.}\ \bibnamefont{Rasio}}}%
   (\bibinfo {year} {2012}),\ \bibinfo {note} {82 pages, 18 figures (reproduced
  with permission of the original authors); to appear in Living Reviews in
  Relativity},\ \Eprint{http://arxiv.org/abs/1204.3858}{arXiv:1204.3858
  [gr-qc]}%
  \bibAnnoteFile{NoStop}{BNSlr}%
%%CITATION = ARXIV:1204.3858;%%
\bibitem{st11}%
  \BibitemOpen
  \bibfield{author}{%
  \bibinfo {author} {\bibfnamefont{M.}~\bibnamefont{Shibata}}\ and\ \bibinfo
  {author} {\bibfnamefont{K.}~\bibnamefont{Taniguchi}},\ }%
  \bibfield{journal}{%
  \bibinfo {journal} {Living Reviews in Relativity}\ }%
  \textbf{\bibinfo {volume} {14}} (\bibinfo {year} {2011}),\
  \url{http://www.livingreviews.org/lrr-2011-6}%
  \bibAnnoteFile{NoStop}{st11}%
\bibitem{Paschalidis:2010dh}%
  \BibitemOpen
  \bibfield{author}{%
  \bibinfo {author} {\bibfnamefont{V.}~\bibnamefont{Paschalidis}}, \bibinfo
  {author} {\bibfnamefont{Z.}~\bibnamefont{Etienne}}, \bibinfo {author}
  {\bibfnamefont{Y.~T.}\ \bibnamefont{Liu}},\ and\ \bibinfo {author}
  {\bibfnamefont{S.~L.}\ \bibnamefont{Shapiro}},\ }%
  \bibfield{journal}{%
  \Doi{10.1103/PhysRevD.83.064002}{\bibinfo {journal} {Phys.Rev.}}\ }%
  \textbf{\bibinfo {volume} {D83}},\ \bibinfo {pages} {064002} (\bibinfo {year}
  {2011}),\ \Eprint{http://arxiv.org/abs/1009.4932}{arXiv:1009.4932
  [astro-ph.HE]}%
  \bibAnnoteFile{NoStop}{Paschalidis:2010dh}%
%%CITATION = ARXIV:1009.4932;%%
\bibitem{PLES2011}%
  \BibitemOpen
  \bibfield{author}{%
  \bibinfo {author} {\bibfnamefont{V.}~\bibnamefont{{Paschalidis}}}, \bibinfo
  {author} {\bibfnamefont{Y.~T.}\ \bibnamefont{{Liu}}}, \bibinfo {author}
  {\bibfnamefont{Z.}~\bibnamefont{{Etienne}}},\ and\ \bibinfo {author}
  {\bibfnamefont{S.~L.}\ \bibnamefont{{Shapiro}}},\ }%
  \bibfield{journal}{%
  \Doi{10.1103/PhysRevD.84.104032}{\bibinfo {journal} {\prd}}\ }%
  \textbf{\bibinfo {volume} {84}},\ \bibinfo {eid} {104032} (\bibinfo {month}
  {Nov.}\ \bibinfo {year} {2011}),\
  \Eprint{http://arxiv.org/abs/1109.5177}{arXiv:1109.5177 [astro-ph.HE]}%
  \bibAnnoteFile{NoStop}{PLES2011}%
\bibitem{Paschalidis:2009zz}%
  \BibitemOpen
  \bibfield{author}{%
  \bibinfo {author} {\bibfnamefont{V.}~\bibnamefont{Paschalidis}}, \bibinfo
  {author} {\bibfnamefont{M.}~\bibnamefont{MacLeod}}, \bibinfo {author}
  {\bibfnamefont{T.~W.}\ \bibnamefont{Baumgarte}},\ and\ \bibinfo {author}
  {\bibfnamefont{S.~L.}\ \bibnamefont{Shapiro}},\ }%
  \bibfield{journal}{%
  \Doi{10.1103/PhysRevD.80.024006}{\bibinfo {journal} {Phys.Rev.}}\ }%
  \textbf{\bibinfo {volume} {D80}},\ \bibinfo {pages} {024006} (\bibinfo {year}
  {2009}),\ \Eprint{http://arxiv.org/abs/0910.5719}{arXiv:0910.5719
  [astro-ph.HE]}%
  \bibAnnoteFile{NoStop}{Paschalidis:2009zz}%
%%CITATION = ARXIV:0910.5719;%%
\bibitem{HMNS}%
  \BibitemOpen
  \bibfield{author}{%
  \bibinfo {author} {\bibfnamefont{T.~W.}\ \bibnamefont{Baumgarte}}, \bibinfo
  {author} {\bibfnamefont{S.~L.}\ \bibnamefont{Shapiro}},\ and\ \bibinfo
  {author} {\bibfnamefont{M.}~\bibnamefont{Shibata}},\ }%
  \bibfield{journal}{%
  \Doi{10.1086/312425}{\bibinfo {journal} {Astrophys.J.}}\ }%
  \textbf{\bibinfo {volume} {528}},\ \bibinfo {pages} {L29} (\bibinfo {year}
  {2000}),\
  \Eprint{http://arxiv.org/abs/astro-ph/9910565}{arXiv:astro-ph/9910565
  [astro-ph]}%
  \bibAnnoteFile{NoStop}{HMNS}%
%%CITATION = ASTRO-PH/9910565;%%
\bibitem{CST1994a}%
  \BibitemOpen
  \bibfield{author}{%
  \bibinfo {author} {\bibfnamefont{G.~B.}\ \bibnamefont{{Cook}}}, \bibinfo
  {author} {\bibfnamefont{S.~L.}\ \bibnamefont{{Shapiro}}},\ and\ \bibinfo
  {author} {\bibfnamefont{S.~A.}\ \bibnamefont{{Teukolsky}}},\ }%
  \bibfield{journal}{%
  \Doi{10.1086/173721}{\bibinfo {journal} {\apj}}\ }%
  \textbf{\bibinfo {volume} {422}},\ \bibinfo {pages} {227} (\bibinfo {month}
  {Feb.}\ \bibinfo {year} {1994})%
  \bibAnnoteFile{NoStop}{CST1994a}%
\bibitem{CST1994b}%
  \BibitemOpen
  \bibfield{author}{%
  \bibinfo {author} {\bibfnamefont{G.~B.}\ \bibnamefont{{Cook}}}, \bibinfo
  {author} {\bibfnamefont{S.~L.}\ \bibnamefont{{Shapiro}}},\ and\ \bibinfo
  {author} {\bibfnamefont{S.~A.}\ \bibnamefont{{Teukolsky}}},\ }%
  \bibfield{journal}{%
  \Doi{10.1086/173934}{\bibinfo {journal} {\apj}}\ }%
  \textbf{\bibinfo {volume} {424}},\ \bibinfo {pages} {823} (\bibinfo {month}
  {Apr.}\ \bibinfo {year} {1994})%
  \bibAnnoteFile{NoStop}{CST1994b}%
\bibitem{ST}%
  \BibitemOpen
  \bibfield{author}{%
  \bibinfo {author} {\bibfnamefont{M.}~\bibnamefont{{Shibata}}}\ and\ \bibinfo
  {author} {\bibfnamefont{K.}~\bibnamefont{{Taniguchi}}},\ }%
  \bibfield{journal}{%
  \Doi{10.1103/PhysRevD.73.064027}{\bibinfo {journal} {\prd}}\ }%
  \textbf{\bibinfo {volume} {73}},\ \bibinfo {pages} {064027} (\bibinfo {month}
  {Mar.}\ \bibinfo {year} {2006})%
  \bibAnnoteFile{NoStop}{ST}%
\bibitem{DLSS2004}%
  \BibitemOpen
  \bibfield{author}{%
  \bibinfo {author} {\bibfnamefont{M.~D.}\ \bibnamefont{{Duez}}}, \bibinfo
  {author} {\bibfnamefont{Y.~T.}\ \bibnamefont{{Liu}}}, \bibinfo {author}
  {\bibfnamefont{S.~L.}\ \bibnamefont{{Shapiro}}},\ and\ \bibinfo {author}
  {\bibfnamefont{B.~C.}\ \bibnamefont{{Stephens}}},\ }%
  \bibfield{journal}{%
  \Doi{10.1103/PhysRevD.69.104030}{\bibinfo {journal} {\prd}}\ }%
  \textbf{\bibinfo {volume} {69}},\ \bibinfo {eid} {104030} (\bibinfo {month}
  {May}\ \bibinfo {year} {2004}),\
  \Eprint{http://arxiv.org/abs/arXiv:astro-ph/0402502}{arXiv:astro-ph/0402502}%
  \bibAnnoteFile{NoStop}{DLSS2004}%
\bibitem{dlsss06a}%
  \BibitemOpen
  \bibfield{author}{%
  \bibinfo {author} {\bibfnamefont{M.~D.}\ \bibnamefont{{Duez}}}, \bibinfo
  {author} {\bibfnamefont{Y.~T.}\ \bibnamefont{{Liu}}}, \bibinfo {author}
  {\bibfnamefont{S.~L.}\ \bibnamefont{{Shapiro}}}, \bibinfo {author}
  {\bibfnamefont{M.}~\bibnamefont{{Shibata}}},\ and\ \bibinfo {author}
  {\bibfnamefont{B.~C.}\ \bibnamefont{{Stephens}}},\ }%
  \bibfield{journal}{%
  \Doi{10.1103/PhysRevLett.96.031101}{\bibinfo {journal} {Physical Review
  Letters}}\ }%
  \textbf{\bibinfo {volume} {96}},\ \bibinfo {pages} {031101} (\bibinfo {month}
  {Jan.}\ \bibinfo {year} {2006})%
  \bibAnnoteFile{NoStop}{dlsss06a}%
\bibitem{SHTA2006}%
  \BibitemOpen
  \bibfield{author}{%
  \bibinfo {author} {\bibfnamefont{M.}~\bibnamefont{{Shibata}}}\ and\ \bibinfo
  {author} {\bibfnamefont{K.}~\bibnamefont{{Taniguchi}}},\ }%
  \bibfield{journal}{%
  \Doi{10.1103/PhysRevD.73.064027}{\bibinfo {journal} {\prd}}\ }%
  \textbf{\bibinfo {volume} {73}},\ \bibinfo {eid} {064027} (\bibinfo {month}
  {Mar.}\ \bibinfo {year} {2006}),\
  \Eprint{http://arxiv.org/abs/arXiv:astro-ph/0603145}{arXiv:astro-ph/0603145}%
  \bibAnnoteFile{NoStop}{SHTA2006}%
\bibitem{Note1}%
  \BibitemOpen
  \bibinfo {note} {Note that neutrinos too carry away angular momentum from the
  system, but according to \cite {Baumgarte:1998sn} neutrino emission is very
  inefficient in decreasing the angular momentum of a HMNS.}%
  \bibAnnoteFile{Stop}{Note1}%
\bibitem{2008PhRvD..78h4033B}%
  \BibitemOpen
  \bibfield{author}{%
  \bibinfo {author} {\bibfnamefont{L.}~\bibnamefont{{Baiotti}}}, \bibinfo
  {author} {\bibfnamefont{B.}~\bibnamefont{{Giacomazzo}}},\ and\ \bibinfo
  {author} {\bibfnamefont{L.}~\bibnamefont{{Rezzolla}}},\ }%
  \bibfield{journal}{%
  \Doi{10.1103/PhysRevD.78.084033}{\bibinfo {journal} {\prd}}\ }%
  \textbf{\bibinfo {volume} {78}},\ \bibinfo {eid} {084033} (\bibinfo {month}
  {Oct.}\ \bibinfo {year} {2008}),\
  \Eprint{http://arxiv.org/abs/0804.0594}{arXiv:0804.0594 [gr-qc]}%
  \bibAnnoteFile{NoStop}{2008PhRvD..78h4033B}%
\bibitem{Rezzolla:2010fd}%
  \BibitemOpen
  \bibfield{author}{%
  \bibinfo {author} {\bibfnamefont{L.}~\bibnamefont{Rezzolla}}, \bibinfo
  {author} {\bibfnamefont{L.}~\bibnamefont{Baiotti}}, \bibinfo {author}
  {\bibfnamefont{B.}~\bibnamefont{Giacomazzo}}, \bibinfo {author}
  {\bibfnamefont{D.}~\bibnamefont{Link}},\ and\ \bibinfo {author}
  {\bibfnamefont{J.~A.}\ \bibnamefont{Font}},\ }%
  \bibfield{journal}{%
  \Doi{10.1088/0264-9381/27/11/114105}{\bibinfo {journal} {Class.Quant.Grav.}}\
  }%
  \textbf{\bibinfo {volume} {27}},\ \bibinfo {pages} {114105} (\bibinfo {year}
  {2010}),\ \Eprint{http://arxiv.org/abs/1001.3074}{arXiv:1001.3074 [gr-qc]}%
  \bibAnnoteFile{NoStop}{Rezzolla:2010fd}%
%%CITATION = ARXIV:1001.3074;%%
\bibitem{2011PhRvL.107e1102S}%
  \BibitemOpen
  \bibfield{author}{%
  \bibinfo {author} {\bibfnamefont{Y.}~\bibnamefont{{Sekiguchi}}}, \bibinfo
  {author} {\bibfnamefont{K.}~\bibnamefont{{Kiuchi}}}, \bibinfo {author}
  {\bibfnamefont{K.}~\bibnamefont{{Kyutoku}}},\ and\ \bibinfo {author}
  {\bibfnamefont{M.}~\bibnamefont{{Shibata}}},\ }%
  \bibfield{journal}{%
  \Doi{10.1103/PhysRevLett.107.051102}{\bibinfo {journal} {Physical Review
  Letters}}\ }%
  \textbf{\bibinfo {volume} {107}},\ \bibinfo {eid} {051102} (\bibinfo {month}
  {Jul.}\ \bibinfo {year} {2011}),\
  \Eprint{http://arxiv.org/abs/1105.2125}{arXiv:1105.2125 [gr-qc]}%
  \bibAnnoteFile{NoStop}{2011PhRvL.107e1102S}%
\bibitem{Shapiro}%
  \BibitemOpen
  \bibfield{author}{%
  \bibinfo {author} {\bibfnamefont{S.~L.}\ \bibnamefont{Shapiro}}\ and\
  \bibinfo {author} {\bibfnamefont{S.~A.}\ \bibnamefont{Teukolsky}},\ }%
  \emph{\bibinfo {title} {Black Holes, White Dwarfs, and Neutron Stars}}\
  (\bibinfo {publisher} {John Willey and Sons},\ \bibinfo {year} {1983})%
  \bibAnnoteFile{NoStop}{Shapiro}%
\bibitem{Rosswog:2003rv}%
  \BibitemOpen
  \bibfield{author}{%
  \bibinfo {author} {\bibfnamefont{S.}~\bibnamefont{Rosswog}}\ and\ \bibinfo
  {author} {\bibfnamefont{M.}~\bibnamefont{Liebendoerfer}},\ }%
  \bibfield{journal}{%
  \Doi{10.1046/j.1365-8711.2003.06579.x}{\bibinfo {journal}
  {Mon.Not.Roy.Astron.Soc.}}\ }%
  \textbf{\bibinfo {volume} {342}},\ \bibinfo {pages} {673} (\bibinfo {year}
  {2003}),\
  \Eprint{http://arxiv.org/abs/astro-ph/0302301}{arXiv:astro-ph/0302301
  [astro-ph]}%
  \bibAnnoteFile{NoStop}{Rosswog:2003rv}%
%%CITATION = ASTRO-PH/0302301;%%
\bibitem{Note2}%
  \BibitemOpen
  \bibinfo {note} {Given the low value of $Y_e \approx 0.1$, i.e., of the mean
  number of electrons per baryon found in NSNS mergers in \cite
  {Rosswog:2003rv} in our estimates here we assume for simplicity that almost
  all baryons are neutrons.}%
  \bibAnnoteFile{Stop}{Note2}%
\bibitem{Magnetar}%
  \BibitemOpen
  \bibfield{author}{%
  \bibinfo {author} {\bibfnamefont{R.~C.}\ \bibnamefont{{Duncan}}}\ and\
  \bibinfo {author} {\bibfnamefont{C.}~\bibnamefont{{Thompson}}},\ }%
  \bibfield{journal}{%
  \Doi{10.1086/186413}{\bibinfo {journal} {\apjl}}\ }%
  \textbf{\bibinfo {volume} {392}},\ \bibinfo {pages} {L9} (\bibinfo {month}
  {Jun.}\ \bibinfo {year} {1992})%
  \bibAnnoteFile{NoStop}{Magnetar}%
\bibitem{Rezzolla:2011da}%
  \BibitemOpen
  \bibfield{author}{%
  \bibinfo {author} {\bibfnamefont{L.}~\bibnamefont{Rezzolla}}, \bibinfo
  {author} {\bibfnamefont{B.}~\bibnamefont{Giacomazzo}}, \bibinfo {author}
  {\bibfnamefont{L.}~\bibnamefont{Baiotti}}, \bibinfo {author}
  {\bibfnamefont{J.}~\bibnamefont{Granot}}, \bibinfo {author}
  {\bibfnamefont{C.}~\bibnamefont{Kouveliotou}}, \emph{et~al.},\ }%
  \bibfield{journal}{%
  \bibinfo {journal} {Astrophys.J.}\ }%
  \textbf{\bibinfo {volume} {732}},\ \bibinfo {pages} {L6} (\bibinfo {year}
  {2011}),\ \Eprint{http://arxiv.org/abs/1101.4298}{arXiv:1101.4298
  [astro-ph.HE]}%
  \bibAnnoteFile{NoStop}{Rezzolla:2011da}%
%%CITATION = ARXIV:1101.4298;%%
\bibitem{eflstb08}%
  \BibitemOpen
  \bibfield{author}{%
  \bibinfo {author} {\bibfnamefont{Z.~B.}\ \bibnamefont{{Etienne}}}, \bibinfo
  {author} {\bibfnamefont{J.~A.}\ \bibnamefont{{Faber}}}, \bibinfo {author}
  {\bibfnamefont{Y.~T.}\ \bibnamefont{{Liu}}}, \bibinfo {author}
  {\bibfnamefont{S.~L.}\ \bibnamefont{{Shapiro}}}, \bibinfo {author}
  {\bibfnamefont{K.}~\bibnamefont{{Taniguchi}}},\ and\ \bibinfo {author}
  {\bibfnamefont{T.~W.}\ \bibnamefont{{Baumgarte}}},\ }%
  \bibfield{journal}{%
  \Doi{10.1103/PhysRevD.77.084002}{\bibinfo {journal} {\prd}}\ }%
  \textbf{\bibinfo {volume} {77}},\ \bibinfo {pages} {084002} (\bibinfo {month}
  {Apr.}\ \bibinfo {year} {2008})%
  \bibAnnoteFile{NoStop}{eflstb08}%
\bibitem{elsb09}%
  \BibitemOpen
  \bibfield{author}{%
  \bibinfo {author} {\bibfnamefont{Z.~B.}\ \bibnamefont{{Etienne}}}, \bibinfo
  {author} {\bibfnamefont{Y.~T.}\ \bibnamefont{{Liu}}}, \bibinfo {author}
  {\bibfnamefont{S.~L.}\ \bibnamefont{{Shapiro}}},\ and\ \bibinfo {author}
  {\bibfnamefont{T.~W.}\ \bibnamefont{{Baumgarte}}},\ }%
  \bibfield{journal}{%
  \Doi{10.1103/PhysRevD.79.044024}{\bibinfo {journal} {\prd}}\ }%
  \textbf{\bibinfo {volume} {79}},\ \bibinfo {pages} {044024} (\bibinfo {month}
  {Feb.}\ \bibinfo {year} {2009})%
  \bibAnnoteFile{NoStop}{elsb09}%
\bibitem{els10}%
  \BibitemOpen
  \bibfield{author}{%
  \bibinfo {author} {\bibfnamefont{Z.~B.}\ \bibnamefont{{Etienne}}}, \bibinfo
  {author} {\bibfnamefont{Y.~T.}\ \bibnamefont{{Liu}}},\ and\ \bibinfo {author}
  {\bibfnamefont{S.~L.}\ \bibnamefont{{Shapiro}}},\ }%
  \bibfield{journal}{%
  \Doi{10.1103/PhysRevD.82.084031}{\bibinfo {journal} {\prd}}\ }%
  \textbf{\bibinfo {volume} {82}},\ \bibinfo {pages} {084031} (\bibinfo {month}
  {Oct.}\ \bibinfo {year} {2010})%
  \bibAnnoteFile{NoStop}{els10}%
\bibitem{APPENDIXPAPER}%
  \BibitemOpen
  \bibfield{author}{%
  \bibinfo {author} {\bibfnamefont{Z.~B.}\ \bibnamefont{{Etienne}}}, \bibinfo
  {author} {\bibfnamefont{V.}~\bibnamefont{{Paschalidis}}}, \bibinfo {author}
  {\bibfnamefont{Y.~T.}\ \bibnamefont{{Liu}}},\ and\ \bibinfo {author}
  {\bibfnamefont{S.~L.}\ \bibnamefont{{Shapiro}}},\ }%
  \bibfield{journal}{%
  \bibinfo {journal} {ArXiv e-prints}}%
   (\bibinfo {month} {Oct.}\ \bibinfo {year} {2011}),\
  \Eprint{http://arxiv.org/abs/1110.4633}{arXiv:1110.4633 [astro-ph.HE]}%
  \bibAnnoteFile{NoStop}{APPENDIXPAPER}%
\bibitem{SN}%
  \BibitemOpen
  \bibfield{author}{%
  \bibinfo {author} {\bibfnamefont{M.}~\bibnamefont{{Shibata}}}\ and\ \bibinfo
  {author} {\bibfnamefont{T.}~\bibnamefont{{Nakamura}}},\ }%
  \bibfield{journal}{%
  \bibinfo {journal} {\prd}\ }%
  \textbf{\bibinfo {volume} {52}},\ \bibinfo {pages} {5428} (\bibinfo {month}
  {Nov.}\ \bibinfo {year} {1995})%
  \bibAnnoteFile{NoStop}{SN}%
\bibitem{BS}%
  \BibitemOpen
  \bibfield{author}{%
  \bibinfo {author} {\bibfnamefont{T.~W.}\ \bibnamefont{{Baumgarte}}}\ and\
  \bibinfo {author} {\bibfnamefont{S.~L.}\ \bibnamefont{{Shapiro}}},\ }%
  \bibfield{journal}{%
  \bibinfo {journal} {\prd}\ }%
  \textbf{\bibinfo {volume} {59}},\ \bibinfo {pages} {024007} (\bibinfo {month}
  {Jan.}\ \bibinfo {year} {1998})%
  \bibAnnoteFile{NoStop}{BS}%
\bibitem{GodGauge}%
  \BibitemOpen
  \bibfield{author}{%
  \bibinfo {author} {\bibfnamefont{J.~R.}\ \bibnamefont{{van Meter}}}, \bibinfo
  {author} {\bibfnamefont{J.~G.}\ \bibnamefont{{Baker}}}, \bibinfo {author}
  {\bibfnamefont{M.}~\bibnamefont{{Koppitz}}},\ and\ \bibinfo {author}
  {\bibfnamefont{D.-I.}\ \bibnamefont{{Choi}}},\ }%
  \bibfield{journal}{%
  \Doi{10.1103/PhysRevD.73.124011}{\bibinfo {journal} {\prd}}\ }%
  \textbf{\bibinfo {volume} {73}},\ \bibinfo {pages} {124011} (\bibinfo {month}
  {Jun.}\ \bibinfo {year} {2006})%
  \bibAnnoteFile{NoStop}{GodGauge}%
\bibitem{web:Lorene}%
  \BibitemOpen
  \bibinfo {note} {{\tt http://www.lorene.obspm.fr/}}%
  \bibAnnoteFile{NoStop}{web:Lorene}%
\bibitem{Cactus}%
  \BibitemOpen
  \bibinfo {note} {{\tt http://www.cactuscode.org/}}%
  \bibAnnoteFile{NoStop}{Cactus}%
\bibitem{Carpet}%
  \BibitemOpen
  \bibfield{author}{%
  \bibinfo {author} {\bibfnamefont{E.}~\bibnamefont{Schnetter}}, \bibinfo
  {author} {\bibfnamefont{S.~H.}\ \bibnamefont{Hawley}},\ and\ \bibinfo
  {author} {\bibfnamefont{I.}~\bibnamefont{Hawke}},\ }%
  \bibfield{journal}{%
  \Doi{10.1088/0264-9381/21/6/014}{\bibinfo {journal} {Class. Quantum Grav.}}\
  }%
  \textbf{\bibinfo {volume} {21}},\ \bibinfo {pages} {1465} (\bibinfo {year}
  {2004}),\
  \Eprint{http://arxiv.org/abs/arXiv:gr-qc/0310042}{arXiv:gr-qc/0310042},\
  \url{http://arxiv.org/abs/gr-qc/0310042}%
  \bibAnnoteFile{NoStop}{Carpet}%
\bibitem{DLSS}%
  \BibitemOpen
  \bibfield{author}{%
  \bibinfo {author} {\bibfnamefont{M.~D.}\ \bibnamefont{{Duez}}}, \bibinfo
  {author} {\bibfnamefont{Y.~T.}\ \bibnamefont{{Liu}}}, \bibinfo {author}
  {\bibfnamefont{S.~L.}\ \bibnamefont{{Shapiro}}},\ and\ \bibinfo {author}
  {\bibfnamefont{B.~C.}\ \bibnamefont{{Stephens}}},\ }%
  \bibfield{journal}{%
  \Doi{10.1103/PhysRevD.72.024028}{\bibinfo {journal} {\prd}}\ }%
  \textbf{\bibinfo {volume} {72}},\ \bibinfo {pages} {024028} (\bibinfo {month}
  {Jul.}\ \bibinfo {year} {2005})%
  \bibAnnoteFile{NoStop}{DLSS}%
\bibitem{PPM}%
  \BibitemOpen
  \bibfield{author}{%
  \bibinfo {author} {\bibfnamefont{P.}~\bibnamefont{{Colella}}}\ and\ \bibinfo
  {author} {\bibfnamefont{P.~R.}\ \bibnamefont{{Woodward}}},\ }%
  \bibfield{journal}{%
  \Doi{10.1016/0021-9991(84)90143-8}{\bibinfo {journal} {Journal of
  Computational Physics}}\ }%
  \textbf{\bibinfo {volume} {54}},\ \bibinfo {pages} {174} (\bibinfo {month}
  {Sep.}\ \bibinfo {year} {1984})%
  \bibAnnoteFile{NoStop}{PPM}%
\bibitem{HLL}%
  \BibitemOpen
  \bibfield{author}{%
  \bibinfo {author} {\bibfnamefont{A.}~\bibnamefont{{Harten}}}, \bibinfo
  {author} {\bibfnamefont{P.}~\bibnamefont{{Lax}}},\ and\ \bibinfo {author}
  {\bibfnamefont{B.}~\bibnamefont{{van Leer}}},\ }%
  \bibfield{journal}{%
  \bibinfo {journal} {SIAM Rev.}\ }%
  \textbf{\bibinfo {volume} {25}},\ \bibinfo {pages} {35} (\bibinfo {year}
  {1983})%
  \bibAnnoteFile{NoStop}{HLL}%
\bibitem{MihalasBook}%
  \BibitemOpen
  \bibfield{author}{%
  \bibinfo {author} {\bibfnamefont{D.}~\bibnamefont{Mihalas}}\ and\ \bibinfo
  {author} {\bibfnamefont{B.~W.}\ \bibnamefont{Mihalas}},\ }%
  \emph{\bibinfo {title} {Foundations of radiation hydrodynamics}}\ (\bibinfo
  {publisher} {Dover Publications},\ \bibinfo {year} {1999})%
  \bibAnnoteFile{NoStop}{MihalasBook}%
\bibitem{CollapseShapiro1996}%
  \BibitemOpen
  \bibfield{author}{%
  \bibinfo {author} {\bibfnamefont{S.~L.}\ \bibnamefont{{Shapiro}}},\ }%
  \bibfield{journal}{%
  \Doi{10.1086/178065}{\bibinfo {journal} {\apj}}\ }%
  \textbf{\bibinfo {volume} {472}},\ \bibinfo {pages} {308} (\bibinfo {month}
  {Nov.}\ \bibinfo {year} {1996})%
  \bibAnnoteFile{NoStop}{CollapseShapiro1996}%
\bibitem{BFarris2008}%
  \BibitemOpen
  \bibfield{author}{%
  \bibinfo {author} {\bibfnamefont{B.~D.}\ \bibnamefont{{Farris}}}, \bibinfo
  {author} {\bibfnamefont{T.~K.}\ \bibnamefont{{Li}}}, \bibinfo {author}
  {\bibfnamefont{Y.~T.}\ \bibnamefont{{Liu}}},\ and\ \bibinfo {author}
  {\bibfnamefont{S.~L.}\ \bibnamefont{{Shapiro}}},\ }%
  \bibfield{journal}{%
  \Doi{10.1103/PhysRevD.78.024023}{\bibinfo {journal} {\prd}}\ }%
  \textbf{\bibinfo {volume} {78}},\ \bibinfo {pages} {024023} (\bibinfo {month}
  {Jul.}\ \bibinfo {year} {2008}),\
  \Eprint{http://arxiv.org/abs/0802.3210}{arXiv:0802.3210}%
  \bibAnnoteFile{NoStop}{BFarris2008}%
\bibitem{ngmz06}%
  \BibitemOpen
  \bibfield{author}{%
  \bibinfo {author} {\bibfnamefont{S.~C.}\ \bibnamefont{{Noble}}}, \bibinfo
  {author} {\bibfnamefont{C.~F.}\ \bibnamefont{{Gammie}}}, \bibinfo {author}
  {\bibfnamefont{J.~C.}\ \bibnamefont{{McKinney}}},\ and\ \bibinfo {author}
  {\bibfnamefont{L.}~\bibnamefont{{Del Zanna}}},\ }%
  \bibfield{journal}{%
  \Doi{10.1086/500349}{\bibinfo {journal} {\apj}}\ }%
  \textbf{\bibinfo {volume} {641}},\ \bibinfo {pages} {626} (\bibinfo {month}
  {Apr.}\ \bibinfo {year} {2006})%
  \bibAnnoteFile{NoStop}{ngmz06}%
\bibitem{harmsolver}%
  \BibitemOpen
  \bibfield{author}{%
  \bibinfo {author} {\bibfnamefont{S.~C.}\ \bibnamefont{{Noble}}}, \bibinfo
  {author} {\bibfnamefont{C.~F.}\ \bibnamefont{{Gammie}}}, \bibinfo {author}
  {\bibfnamefont{J.~C.}\ \bibnamefont{{McKinney}}},\ and\ \bibinfo {author}
  {\bibfnamefont{L.}~\bibnamefont{{Del Zanna}}}}%
   (\bibinfo {year} {2006}),\ \bibinfo {note} {publicly available on
  http://rainman.astro.illinois.edu/codelib/}%
  \bibAnnoteFile{NoStop}{harmsolver}%
\bibitem{bs2011}%
  \BibitemOpen
  \bibfield{author}{%
  \bibinfo {author} {\bibfnamefont{K.}~\bibnamefont{{Beckwith}}}\ and\ \bibinfo
  {author} {\bibfnamefont{J.~M.}\ \bibnamefont{{Stone}}},\ }%
  \bibfield{journal}{%
  \Doi{10.1088/0067-0049/193/1/6}{\bibinfo {journal} {\apjs}}\ }%
  \textbf{\bibinfo {volume} {193}},\ \bibinfo {pages} {6} (\bibinfo {month}
  {Mar.}\ \bibinfo {year} {2011})%
  \bibAnnoteFile{NoStop}{bs2011}%
\bibitem{Etienne:2011ea}%
  \BibitemOpen
  \bibfield{author}{%
  \bibinfo {author} {\bibfnamefont{Z.~B.}\ \bibnamefont{Etienne}}, \bibinfo
  {author} {\bibfnamefont{Y.~T.}\ \bibnamefont{Liu}}, \bibinfo {author}
  {\bibfnamefont{V.}~\bibnamefont{Paschalidis}},\ and\ \bibinfo {author}
  {\bibfnamefont{S.~L.}\ \bibnamefont{Shapiro}},\ }%
  \bibfield{journal}{%
  \Doi{10.1103/PhysRevD.85.064029}{\bibinfo {journal} {Phys.Rev.}}\ }%
  \textbf{\bibinfo {volume} {D85}},\ \bibinfo {pages} {064029} (\bibinfo {year}
  {2012}),\ \bibinfo {note} {30 pages, 26 figures, 3 tables, submitted to PRD,
  updated references},\ \Eprint{http://arxiv.org/abs/1112.0568}{arXiv:1112.0568
  [astro-ph.HE]}%
  \bibAnnoteFile{NoStop}{Etienne:2011ea}%
%%CITATION = ARXIV:1112.0568;%%
\bibitem{bm09}%
  \BibitemOpen
  \bibfield{author}{%
  \bibinfo {author} {\bibfnamefont{M.}~\bibnamefont{{Boyle}}}\ and\ \bibinfo
  {author} {\bibfnamefont{A.~H.}\ \bibnamefont{{Mrou{\'e}}}},\ }%
  \bibfield{journal}{%
  \Doi{10.1103/PhysRevD.80.124045}{\bibinfo {journal} {\prd}}\ }%
  \textbf{\bibinfo {volume} {80}},\ \bibinfo {pages} {124045} (\bibinfo {month}
  {Dec.}\ \bibinfo {year} {2009})%
  \bibAnnoteFile{NoStop}{bm09}%
\bibitem{rant08}%
  \BibitemOpen
  \bibfield{author}{%
  \bibinfo {author} {\bibfnamefont{M.}~\bibnamefont{{Ruiz}}}, \bibinfo {author}
  {\bibfnamefont{M.}~\bibnamefont{{Alcubierre}}}, \bibinfo {author}
  {\bibfnamefont{D.}~\bibnamefont{{N{\'u}{\~n}ez}}},\ and\ \bibinfo {author}
  {\bibfnamefont{R.}~\bibnamefont{{Takahashi}}},\ }%
  \bibfield{journal}{%
  \Doi{10.1007/s10714-007-0570-8}{\bibinfo {journal} {General Relativity and
  Gravitation}}\ }%
  \textbf{\bibinfo {volume} {40}},\ \bibinfo {pages} {1705} (\bibinfo {month}
  {Aug.}\ \bibinfo {year} {2008})%
  \bibAnnoteFile{NoStop}{rant08}%
\bibitem{Note3}%
  \BibitemOpen
  \bibinfo {note} {The corresponding plots in case B1 are similar and therefore
  not shown.}%
  \bibAnnoteFile{Stop}{Note3}%
\bibitem{Baumgarte:1998sn}%
  \BibitemOpen
  \bibfield{author}{%
  \bibinfo {author} {\bibfnamefont{T.~W.}\ \bibnamefont{Baumgarte}}\ and\
  \bibinfo {author} {\bibfnamefont{S.~L.}\ \bibnamefont{Shapiro}},\ }%
  \bibfield{journal}{%
  \Doi{10.1086/306067}{\bibinfo {journal} {Astrophys.J.}}\ }%
  \textbf{\bibinfo {volume} {504}},\ \bibinfo {pages} {431} (\bibinfo {year}
  {1998}),\
  \Eprint{http://arxiv.org/abs/astro-ph/9801294}{arXiv:astro-ph/9801294
  [astro-ph]}%
  \bibAnnoteFile{NoStop}{Baumgarte:1998sn}%
%%CITATION = ASTRO-PH/9801294;%%
\bibitem{Sekiguchi:2010fh}%
  \BibitemOpen
  \bibfield{author}{%
  \bibinfo {author} {\bibfnamefont{Y.}~\bibnamefont{Sekiguchi}}}%
   (\bibinfo {year} {2010}),\
  \Eprint{http://arxiv.org/abs/1009.3358}{arXiv:1009.3358 [astro-ph.HE]}%
  \bibAnnoteFile{NoStop}{Sekiguchi:2010fh}%
%%CITATION = ARXIV:1009.3358;%%
\end{thebibliography}%

\end{document}